\documentstyle[12pt,epsf]{article}
\catcode`@=11 

\renewcommand{\theequation}{\arabic{equation}}

\newtoks\@stequation
 
\def\subequations{\refstepcounter{equation}%
  \edef\@savedequation{\the\c@equation}%
  \@stequation=\expandafter{\theequation}
  \edef\@savedtheequation{\the\@stequation}
  \edef\oldtheequation{\theequation}%
  \setcounter{equation}{0}%
  \def\theequation{\oldtheequation\alph{equation}}}
 
\def\endsubequations{%
  \ifnum\c@equation < 2 \@warning{Only \the\c@equation\space subequation
    used in equation \@savedequation}\fi
  \setcounter{equation}{\@savedequation}%
  \@stequation=\expandafter{\@savedtheequation}%
  \edef\theequation{\the\@stequation}%
  \global\@ignoretrue}
 
\def\eqnarray{\stepcounter{equation}\let\@currentlabel\theequation
\global\@eqnswtrue\m@th
\global\@eqcnt\z@\tabskip\@centering\let\\\@eqncr
$$\halign to\displaywidth\bgroup\@eqnsel\hskip\@centering
     $\displaystyle\tabskip\z@{##}$&\global\@eqcnt\@ne
      \hfil$\;{##}\;$\hfil
     &\global\@eqcnt\tw@ $\displaystyle\tabskip\z@{##}$\hfil
   \tabskip\@centering&\llap{##}\tabskip\z@\cr}
\newcommand{\DS}{\displaystyle}
\newcommand{\FS}{\footnotesize}
\def\simgt{\mathop{>}\limits_{\displaystyle{\sim}}}

\begin{document}

\begin{flushright}
OCHA-PP-130 \\
AJC-HEP-32 \\
December 16, 1999\\
Revised January 30, 2000
\end{flushright}

\vfil

\begin{center}
\Large
 Production of Scalar Higgs and Pseudoscalar Higgs \\
 in Multi-Higgs Doublet Models 
 at $\gamma\gamma$ Colliders
\vfil

\normalsize \rm 
 Eri {\sc Asakawa}, \\[3mm]
 {\em Graduate School of Humanities and Sciences, 
 Ochanomizu University \\
      1-1 Otsuka 2-chome, Bunkyo, Tokyo 112-8610, JAPAN }\\
 {\it e-mail: \verb+g9870401@edu.cc.ocha.ac.jp+}\\[7mm]

 Jun-ichi {\sc Kamoshita} and 
 Akio {\sc Sugamoto}\\[3mm]
 {\em Department of Physics, Ochanomizu University \\
      1-1 Otsuka 2-chome, Bunkyo, Tokyo 112-8610, JAPAN }\\[7mm]
and \\[7mm]
 Isamu {\sc Watanabe}\\[3mm]
 {\em Akita Keizaihoka University Junior College \\
      46-1 Morisawa, Sakura, Shimokita-te, Akita 010-8515, JAPAN}\\
 {\it e-mail: \verb+isamu@akeihou-u.ac.jp+}
\vfil
{\bf \sc Abstract}
\end{center}

 We present the effects of heavy CP-even ($H$) and CP-odd ($A$) 
Higgs bosons on the production cross section of the process 
$\gamma\gamma$ $\rightarrow t\overline{t}$ at the energy around the 
mass poles of the Higgs bosons.  
 It is found that the interference between $H$ and $A$ with 
the small mass gap, as well as the ones between Higgs bosons and 
the continuum, contributes to the cross section, if the photon beams 
are polarized and if we observe the helicity of the top quarks.  
 It is demonstrated in the framework of the minimal supersymmetric 
extension of the standard model that the $H$ and $A$ contributions 
can be sizable at future $\gamma\gamma$ colliders for small value of 
$\tan\beta$.  
 The methods to measure the CP-parity of the Higgs boson are also 
presented.  
 The statistical significances of detecting the Higgs signals and 
measuring the Higgs CP-parity are evaluated.

\newpage
\section{Introduction}
\label{S1}

 Search for the Higgs bosons and the precise measurements on their 
properties, such as the masses, the decay widths and the decay 
branching ratios, are the most important subjects to study the 
mechanism of the electroweak symmetry breaking.  
 In the standard model (SM) of the particle physics, only one 
physical neutral Higgs boson appears.  
 On the other hand, models with multiple Higgs doublets have 
CP-even and odd neutral Higgs bosons and charged Higgs bosons, 
if CP is a good symmetry.  

 It is of great interest to examine how we are able to observe these 
Higgs bosons, and how their signals look like.  
 A $\gamma\gamma$ option [\ref{ggoriginal}] of the future linear 
$e^+ e^-$ colliders gives an ideal place to look for such Higgs 
signals [\ref{ggcol}].  
 Neutral Higgs bosons can be produced via loop diagrams of charged 
particles.  
 In the SM case, if the Higgs boson is lighter than about 140 GeV, 
it can be detected and its two-photon decay width can be measured 
accurately by looking for the $b\overline{b}$ pair decay mode of 
the Higgs boson [\ref{OTW}].  
 When the Higgs boson is heavy in the SM, the process $\gamma\gamma 
\rightarrow ZZ$ is useful to detect its signal [\ref{Jikia}], since 
there is no tree-level background in this process.  

 It is also expected that in the extended models of the Higgs sector, 
the detection of a light Higgs boson may be performed just like in 
the SM, as far as it decays dominantly into a $b\overline{b}$ pair.  
 On the other hand, it may be ineffective to look for a heavy CP-even 
Higgs boson ($H$) in the $ZZ$ decay mode, since the branching 
fraction may be suppressed as often happens in the supersymmetric 
extensions of the SM.  
 Furthermore, a CP-odd Higgs boson ($A$) does not have the $ZZ$ decay 
mode.  
 Then, the $t\overline{t}$ decay mode can be used, 
if the Higgs bosons in interest are sufficiently heavy 
and have substantial branching fractions for the $t \overline{t}$ 
decay. 
 In this case, both $H$ and $A$ bosons may contribute to 
the process $\gamma\gamma 
\rightarrow t\overline{t}$ around their mass poles.  

 It is also notable that $H$ and $A$ bosons acquire masses with 
similar magnitudes for some appropriate values of parameters in the 
Higgs sector.  
 The amplitudes of $\gamma\gamma \rightarrow H \rightarrow 
t\overline{t}$ and $\gamma\gamma \rightarrow A \rightarrow 
t\overline{t}$ can interfere, if the mass difference between $H$ and 
$A$ bosons is smaller than or in the same order of the decay widths of 
these bosons, and if the helicities of the initial and final particles 
are properly selected.  
 Additional interferences between these Higgs production amplitudes 
and the continuum one should also be taken into account, if 
the resonant and the continuum amplitudes have comparable magnitudes 
near the resonance.  
 This actually occurs in the case of heavy Higgs bosons, since the 
total decay widths in this case are large enough to reduce the peak 
magnitudes of the Higgs production amplitudes near the mass poles.  

 In this paper, we study production and decay of heavy Higgs bosons 
of both CP-parity in the process $\gamma\gamma \rightarrow 
t\overline{t}$.  
 We also discuss the feasibility of detecting heavy Higgs bosons at 
future $\gamma\gamma$ colliders by using the minimal supersymmetric 
extension of the SM (MSSM) as an example.  
 We pay special attention to the interference effects between the 
variously contributing helicity amplitudes in the process.  
 Focusing on the interference effects we propose a new method to 
measure the CP-parity of Higgs bosons.

 The organization of this paper is as follows.  
 In the next section, we study the interference of $H$ and $A$ bosons 
in the process $\gamma\gamma \rightarrow t\overline{t}$.  
 The helicity amplitudes are calculated in section \ref{S3}, and the 
numerical results are given in section \ref{S4}.  
 We give the conclusions in the last section.  

 Preliminary results of this work have been reported at the 
``LCWS99"~[\ref{asa}].

\section{Interference between Higgs-exchange Amplitudes}
\label{S2}

\begin{figure}[ht]

\section*{Figures}

\vfil

\begin{center}
\hspace*{15mm}
\epsfxsize=7cm
\epsffile{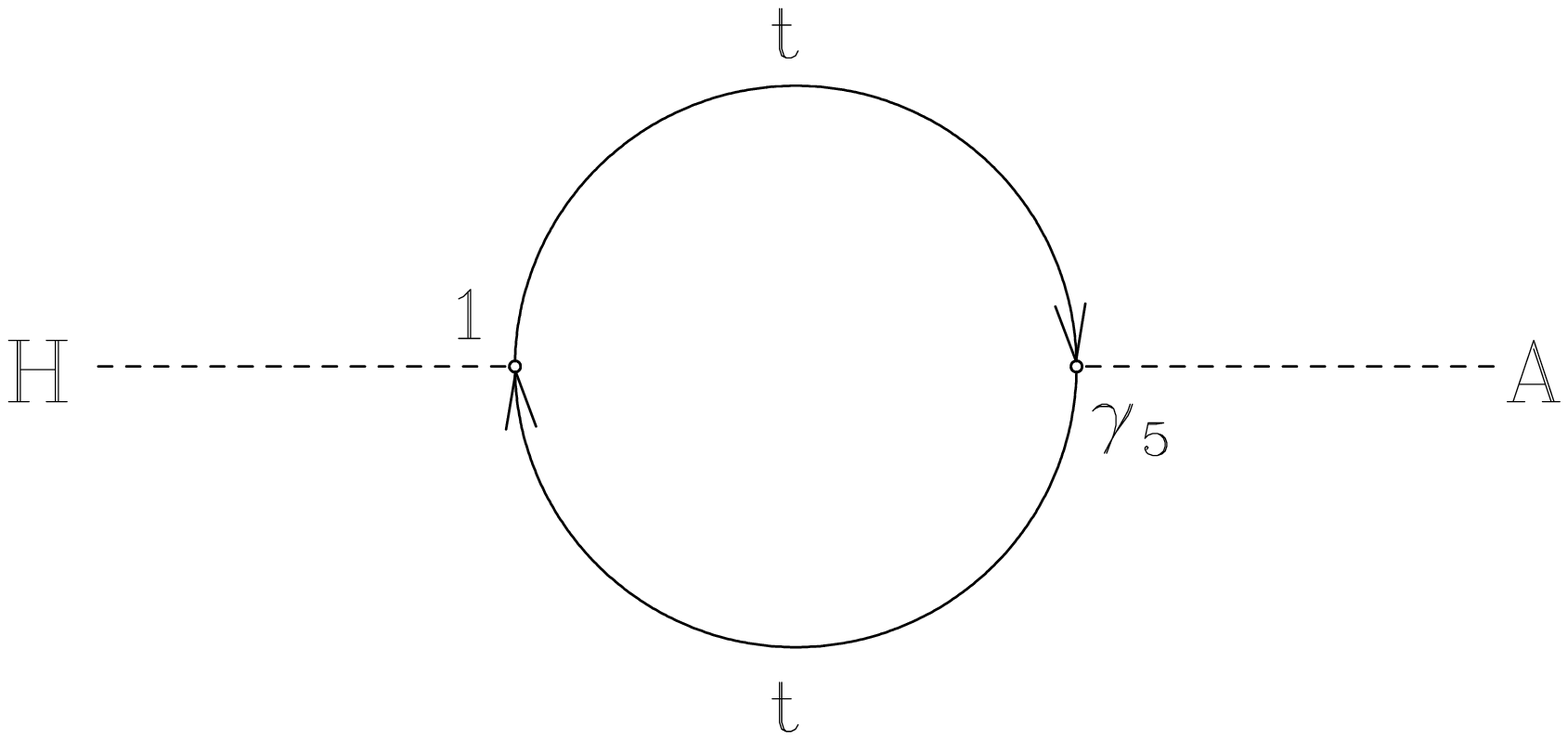}
\vspace*{-10mm}
\caption[Top-loop diagram of $H$-$A$ transition]{\small 
 The top-loop amplitude that connects $H$ and $A$.}
\label{trace}
\end{center}

\vfil

\begin{center}
\hspace*{15mm}
\epsfxsize=10cm
\epsffile{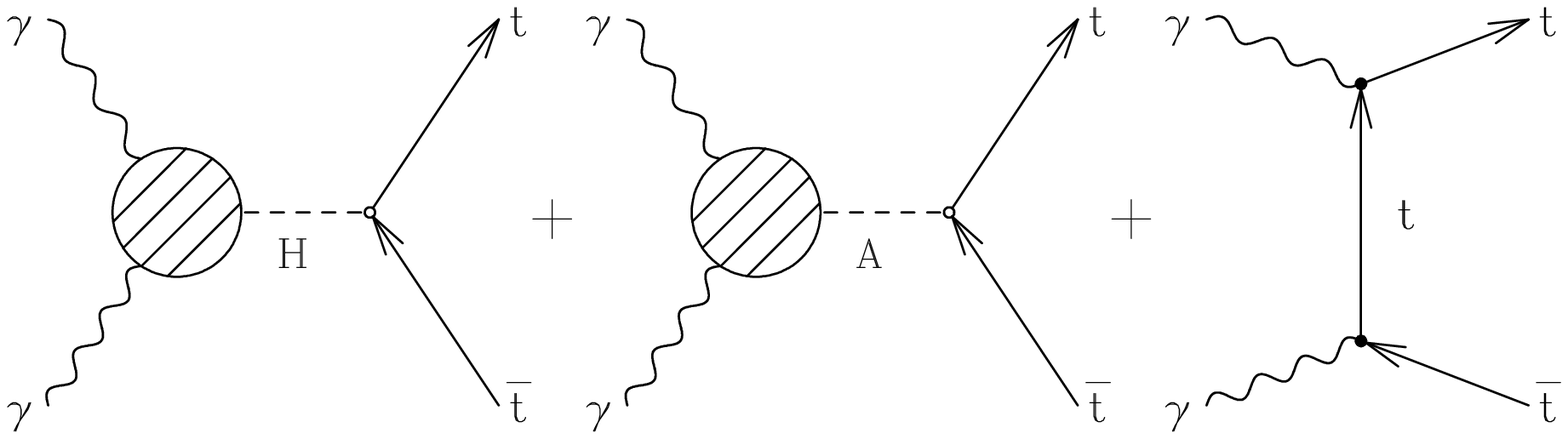}
\vspace*{-15mm}
\caption[Diagrams of $\gamma\gamma \rightarrow t\overline{t}$]%
{\small 
 The dominant diagrams of the process $\gamma\gamma$ $\rightarrow 
t\overline{t}$ at around the mass poles of $H$ and $A$ bosons.  
 The u-channel tree diagram is omitted in the figure.  }
\label{diagrams}
\end{center}
\end{figure}

 One may naively expect that the interference between the 
$H$-exchange and the $A$-exchange amplitudes must vanish since $H$ 
and $A$ bosons have opposite CP parities.  
 In fact, the top-loop amplitude that connects $H$ and $A$ bosons 
(Fig.~\ref{trace}) can easily be evaluated to be zero by using the 
trace technique of undergraduate exercises:  
\begin{equation}
\mbox{Tr} [ \gamma_5 ( p\hspace{-0.5em}/_t + m_t )
 ( - p\hspace{-0.5em}/_{\overline{t}} + m_t ) ] \ = \ 0.  
\label{eq-trace}
\end{equation}
 Here we expressed the four-momenta of top and anti-top quarks as 
$p_t$ and $p_{\bar{t}}$, respectively.  
 Taking the trace of the top-loop means summing up all the different 
helicities of the $t$ and $\overline{t}$.  
 We can evade this summation, however, by selecting helicity states 
of the decaying $t$ and $\overline{t}$ from $H$ and $A$, since the 
helicity of a top quark can be determined statistically due to the 
decay angle dependence of $t \rightarrow b W^+$ [\ref{HMW}], that is, 
we can measure the helicities of $t$ and $\overline{t}$ because 
their decays into $b$ and $W$ are parity-violating.  
 It is in contrast to the production of lighter quarks which mostly 
form the low-lying pseudoscalar mesons.  

 Denoting the helicities of $t$ and $\overline{t}$ as $\lambda$ and 
$\overline{\lambda}$, respectively in the $t\overline{t}$ c.m.\ 
frame, there are two ways of forming a 
$t\overline{t}$ system from a decay of a spin-zero state 
$\langle \lambda\overline{\lambda}|$; {\it i.e.}, $\lambda$ 
$= \overline{\lambda}$ $= L$ ($\langle LL|$) and $\lambda$ 
$= \overline{\lambda}$ $= R$ ($\langle RR|$).  
 The CP transformation $\mbox{\sf CP}$ interchanges $t_L$ and 
$\overline{t}_R$, and $t_R$ and $\overline{t}_L$, with an additional 
minus sign due to the parity difference between the particle and the 
anti-particle states:  
\begin{subequations}
\begin{eqnarray}
\mbox{\sf CP} \langle LL| & = & - \langle RR| \ , \\
\mbox{\sf CP} \langle RR| & = & - \langle LL| \ .  
\end{eqnarray}
\label{parity}
\end{subequations}
\hspace*{-0.55em}
 The CP eigenstates are 45$^\circ$ mixtures of $\langle LL|$ and 
$\langle RR|$:  
\begin{subequations}
\begin{eqnarray}
\mbox{\sf CP} (\langle LL| - \langle RR|) & = & 
 + (\langle LL| - \langle RR|) \ , 
\label{3a} \\
\mbox{\sf CP} (\langle LL| + \langle RR|) & = & 
 - (\langle LL| + \langle RR|) \ .  
\label{3b}  
\end{eqnarray}
\label{eigen}
\end{subequations}
\hspace*{-0.55em}
 The state (\ref{3a}) couples to the CP-even boson $H$, and the state 
(\ref{3b}) couples to the CP-odd boson $A$.  
 Therefore, the states $\langle LL|$ and $\langle RR|$ are relevant 
to the difference and the sum of the $A$-exchange and the $H$-exchange 
amplitudes, respectively.  

 A similar argument can be repeated to the production amplitudes of 
$\gamma\gamma \rightarrow H/A$.  
 In the frame of the linear polarization of initial photons, $H$ and 
$A$ states can be produced from collisions of the parallel and the 
perpendicularly polarized photons, respectively [\ref{linearHA}].  
 Though they are the CP eigenstates, the helicity eigenstates 
$|++ \rangle$ and $|-- \rangle$ are mixtures of them:  
\begin{subequations}
\begin{eqnarray}
|++ \rangle & = & 
[(|xx \rangle - |yy \rangle) 
 + i (|xy \rangle + |yx \rangle)]/2 \ , \\
|-- \rangle & = & 
[(|xx \rangle - |yy \rangle) 
 - i (|xy \rangle + |yx \rangle)]/2 \ , 
\end{eqnarray}
\label{photon-eigen}
\hspace*{-0.9em}
\end{subequations}
where we express the helicities of two photons in the kets in the 
left-hand-side of the equations, while the orientations of the 
linear polarizations of the photons are given in the right-hand-side 
when the momenta of the photons are along the $z$-axis direction.  
 Then, we can expect the $H$--$A$ interference with the circularly 
polarized initial photons.  

 A $\gamma\gamma$ collider option [\ref{ggoriginal},\ref{ggcol}] of 
the future linear colliders is an ideal place to look for the above 
interference in the process $\gamma\gamma \rightarrow H/A \rightarrow 
t\overline{t}$, since the circular polarization of the photon beams 
are naturally realized at the $\gamma\gamma$ colliders.

\section{Helicity Amplitudes}
\label{S3}

 As an example of the Higgs sector with multiple doublets, we study 
the MSSM having two Higgs doublets.  
 There, we are able to demonstrate the definite numerical estimations 
with fixing a few MSSM parameters, without loosing the generality 
and the essence which are applicable for more complicated models.  
 We have three neutral Higgs boson states in the MSSM, the light 
Higgs $h$, the heavy Higgs $H$ and the pseudoscalar Higgs $A$.  
 The former two are CP-even, and the last one is CP-odd.  
 As in the literature, we parameterize the Higgs sector by two 
parameters, the mass of $A$, $m_A$, and the ratio of the vacuum 
expectation values of two Higgs doublets, $\tan\beta$.  
 In the MSSM, if $A$ is lighter than the $Z$ boson, the mass of $h$, 
$m_h$, is similar to $m_A$.  
 On the other hand, as $m_A$ becomes heavier, the mass of $H$, $m_H$ 
approaches to $m_A$, while $m_h$ remains below 150 GeV.  
 It is also noted that the current experimental lower bounds of $m_A$ 
and $m_h$ are 84.5 GeV and 84.3 GeV, respectively [\ref{mlimit}].  

 Assuming $m_A$ $\gg m_Z$, a possibility arises for observing the 
interference between $H$ and $A$ bosons.  
 The process to be focused on is $\gamma\gamma \rightarrow 
t\overline{t}$ with the fixed helicities of $t$ and $\overline{t}$, 
and with the circular polarization of the initial photons.  
 Both $H$ and $A$ bosons can contribute to the amplitudes around their 
mass poles as well as the continuum amplitudes 
(Fig.~\ref{diagrams}).  
 We denote these amplitudes separately as, 
\begin{equation}
{\cal M}_{\gamma\gamma \rightarrow t\overline{t}}%
        ^{\Lambda \overline{\Lambda} \lambda \overline{\lambda}} 
 \simeq 
  {\cal M}_H^{\Lambda \overline{\Lambda} \lambda \overline{\lambda}} 
+ {\cal M}_A^{\Lambda \overline{\Lambda} \lambda \overline{\lambda}}
+ {\cal M}_{\mbox{\tiny tree}}%
          ^{\Lambda \overline{\Lambda} \lambda \overline{\lambda}} \ , 
\label{amp}
\end{equation}
where the subscripts $H$, $A$ and `tree' mean the $H$-exchange, 
the $A$-exchange and the tree diagrams, respectively.  
 The superscripts $\Lambda$ and $\overline{\Lambda}$ denote the 
initial photon helicities, $\lambda$ and $\overline{\lambda}$ denote 
$t$ and $\overline{t}$ helicities, respectively.  

 The Higgs-exchange amplitudes 
${\cal M}_H^{\Lambda \overline{\Lambda} \lambda \overline{\lambda}}$ 
and 
${\cal M}_A^{\Lambda \overline{\Lambda} \lambda \overline{\lambda}}$ 
are then given by the simple multiplication of the Higgs-$\gamma\gamma$ 
vertex function ${\cal A}_{H,A}^{\Lambda \overline{\Lambda}}$, the 
Higgs propagator ${\cal B}_{H,A}$ and the decay part 
${\cal C}_{H,A}^{\lambda \overline{\lambda}}$ of $H,A$ 
$\rightarrow t\overline{t}$:  
\begin{subequations}
\begin{eqnarray}
{\cal M}_H^{\Lambda \overline{\Lambda} \lambda \overline{\lambda}} 
 & = & 
{\cal A}_H^{\Lambda \overline{\Lambda}} {\cal B}_H 
{\cal C}_H^{\lambda \overline{\lambda}} \ , \\
{\cal M}_A^{\Lambda \overline{\Lambda} \lambda \overline{\lambda}} 
 & = & 
{\cal A}_A^{\Lambda \overline{\Lambda}} {\cal B}_A 
{\cal C}_A^{\lambda \overline{\lambda}} \ .  
\end{eqnarray}
\label{parts}
\end{subequations}

\subsection{Higgs--Photon Couplings}

 In the one-loop approximation, $W$ bosons, quarks, charged leptons, 
squarks, charged sleptons, charged Higgs bosons and charginos 
contribute to the $H\gamma\gamma$ vertex function 
${\cal A}_H^{\Lambda \overline{\Lambda}}$.  
 Only quarks, charged leptons and charginos contribute to 
${\cal A}_A^{\Lambda \overline{\Lambda}}$.  
 The dominant contributors are the top quark and the $W$ boson (only 
in the $H$ case) for small values of $\tan\beta$, if their 
superpartners are heavy enough comparing with Higgs bosons.  
 The absolute values of ${\cal A}_{H,A}^{\Lambda \overline{\Lambda}}$ 
can be derived from the formulae of the Higgs partial decay widths 
into two photons $\Gamma(H,A \rightarrow \gamma\gamma)$, which are 
found in the literature [\ref{Hunter}].  
 It is essential to know not only the absolute values but also the 
relative phases of these amplitudes, since we wish to compute the 
interference.  
 The formulae of $\Gamma(H,A \rightarrow \gamma\gamma)$ in the 
literature keep the relative phases of the relevant loops only in 
each amplitude of $H$ and $A$ bosons, but the relative phases between 
them are missing, since only the absolute values of the respective 
amplitudes have been necessary to obtain the corresponding decay 
widths.  
 Therefore, we have calculated the relative phase of the $H$-exchange 
and the $A$-exchange amplitudes by using the top-quark contribution.  
 We have found that ${\cal A}_A^{\Lambda \overline{\Lambda}}$ changes 
its sign according to the helicities of the initial photons, while 
${\cal A}_H^{\Lambda \overline{\Lambda}}$ is independent of the 
photons helicities:  
\begin{subequations}
\begin{eqnarray}
{\cal A}_H^{\Lambda \overline{\Lambda}} \ = & 
-\frac{\DS \alpha_{\mbox{\tiny QED}}g}{\DS 8\pi} 
\frac{\DS m_H^2}{\DS m_W}
\DS \sum_i I^i_H \qquad & 
\mbox{(for $\Lambda \overline{\Lambda}$ = $\pm \pm$)}, 
\label{loop-h} \\
{\cal A}_A^{\Lambda \overline{\Lambda}} \ = & 
\pm i \frac{\DS \alpha_{\mbox{\tiny QED}}g}{\DS 8\pi} 
\frac{\DS m_A^2}{\DS m_W}
\DS \sum_i I^i_A \qquad & 
\mbox{(for $\Lambda \overline{\Lambda}$ = $\pm \pm$)}, 
\label{loop-a} \\
{\cal A}_H^{\Lambda \overline{\Lambda}} \ = & \ 
{\cal A}_A^{\Lambda \overline{\Lambda}} = 0 
\qquad \qquad \qquad \quad & 
\mbox{(for $\Lambda \overline{\Lambda}$ = $\pm \mp$)}, 
\end{eqnarray}
\label{loops}
\end{subequations}
\hspace*{-0.5em}
where $g$ is the weak coupling constant, $m_W$ is the $W$ boson mass 
and the functions $I^i_{H,A}$ can be found in eq.~C.4 of 
Ref.~\ref{Hunter}.  
 The photon-helicity dependent sign appears only in 
eq.~(\ref{loop-a}), which is consistent with our previous observation 
in eq.~(\ref{photon-eigen}).

\subsection{Higgs Propagators}

 It is instructive to study the energy dependence of the propagators 
${\cal B}_{H,A}$ in the complex plane.  
 A propagator with the decay width in its denominator draws a circle 
in the complex plane as its four-momentum squared $q^2$ increases:  
\begin{equation}
{\cal B}/i 
= \frac{i}{q^2 - m^2 + i m \Gamma} 
= \frac{1}{2 m \Gamma} 
\Biggm[ 1 + 
\exp \Big( 2i \tan^{-1} \frac{q^2 - m^2}{m \Gamma} \Big) \Biggm] \ , 
\label{circle}
\end{equation}
where $m$ and $\Gamma$ are the mass and the decay width of the 
propagating particle, respectively.  
 The dominant energy dependence of the Higgs-exchange amplitudes 
${\cal M}_{H,A}$ around the mass poles comes from this circular 
motion in the complex plane, if the energy is far from the thresholds 
of the loop particles.

\subsection{Higgs--Top Couplings}

 The decay parts ${\cal C}_{H,A}^{\lambda \overline{\lambda}}$ are 
simply evaluated in the tree approximation:  
\begin{subequations}
\begin{eqnarray}
{\cal C}_H^{\lambda \overline{\lambda}} \ = & 
\mp \frac{\DS g m_t}{\DS m_W} \frac{\DS \sin\alpha}{\DS \sin\beta} 
E_t \beta_t \qquad & 
(\mbox{for $\lambda \overline{\lambda}$ = ${}^R_L {}^R_L$}), 
\label{vertex-h} \\
{\cal C}_A^{\lambda \overline{\lambda}} \ = & 
- i \frac{\DS g m_t}{\DS m_W} \cot\beta E_t \qquad \quad & 
(\mbox{for $\lambda \overline{\lambda}$ = ${}^R_L {}^R_L$}), 
\label{vertex-a} \\
{\cal C}_H^{\lambda \overline{\lambda}} \ = & \ 
{\cal C}_A^{\lambda \overline{\lambda}} = 0 
\qquad \qquad \qquad & 
(\mbox{for $\lambda \overline{\lambda}$ = ${}^R_L {}^L_R$}), 
\end{eqnarray}
\end{subequations}
where $\alpha$ is the mixing angle of the two neutral CP-even Higgs 
bosons, $E_t$ and $\beta_t$ are the energy and the velocity of 
the decaying top quark in the center of mass frame, respectively.  
 Note here that the top-quark-helicity dependent sign appears only in 
the $H$ amplitude of eq.~(\ref{vertex-h}), while it does not appear 
in the $A$ amplitude of eq.~(\ref{vertex-a}).  
 They are consistent with our previous observation in 
eq.~(\ref{eigen}).

\subsection{Tree Amplitudes}

 In Table~\ref{tree}, ${\cal M}_{\mbox{\tiny tree}}%
^{\Lambda \overline{\Lambda} 
\lambda \overline{\lambda}}$ for each helicity combinations of 
external particles are summarized.  
 Here $Q_t$ and $\theta_t$ are the electric charge and the scattering 
angle of the top quark, respectively.  
 The factor $m_t/E_t$ appearing in the amplitudes with $\lambda$ 
$= \overline{\lambda}$ is a consequence of the helicity conservation 
law in the massless limit of the fermion.  
 Note that the CP transformation interchanges $\Lambda$ and 
$-\Lambda$, $\overline{\Lambda}$ and $-\overline{\Lambda}$, $t_R$ and 
$\overline{t}_L$, $t_L$ and $\overline{t}_R$, and $\cos\theta_t$ and 
$-\cos\theta_t$, each other, with an additional overall minus sign:  
{\it eg.}, ${\cal M}_{\mbox{\tiny tree}}^{++RR}$ 
$= - {\cal M}_{\mbox{\tiny tree}}^{--LL}$.  
 If the circular polarizations of the initial photon beams are purely 
tuned to 100{\%}, there are only $t_R \overline{t}_R$ and 
$t_L \overline{t}_L$ productions.  
 In the case of $\Lambda = \overline{\Lambda} = +$, the absolute 
value of the tree amplitude of $t_R \overline{t}_R$ is much greater 
than that of $t_L \overline{t}_L$ at high energies, due to difference 
of the factor $1+\beta_t$ and $1-\beta_t$.

\section{Numerical Estimates}
\label{S4}

 For our numerical estimates, the following input parameters are 
adopted:  
$m_A$ = 400 GeV, 
$m_W$ = 80.41 GeV, 
$m_t$ = 175.0 GeV, 
the bottom quark pole mass $m_b$ = 4.6 GeV, 
the charm quark pole mass $m_c$ = 1.4 GeV, 
the tau lepton mass $m_\tau$ = 1.777 GeV, 
the weak mixing angle $\sin^2\theta_W$ = 0.2312, 
the QCD coupling constant at the $Z$ boson mass scale 
$\alpha_s(m_Z)$ = 0.117.  
 Since we don't want an undesirable complication, we have chosen 
the sfermion mass scale $m_{\mbox{\tiny SUSY}}$ = 1 TeV, 
the SU(2) gaugino mass parameter $M_2$ = 500 GeV and 
the higgsino mixing mass parameter $\mu$ = $-500$ GeV, 
which result in heavy supersymmetric particles.  
 Thus, in the two-photon decay amplitudes of the Higgs bosons, the 
contributions of the supersymmetric loops are suppressed.  
 The masses, the decay widths and the decay branching ratios of the 
Higgs bosons are evaluated by the program HDECAY~[\ref{HDECAY}] with 
choosing a value of $\tan\beta$.  
 With the above parameters and the value of $\tan\beta$ between 1 and 
20, we found that no real production of the new particles other than 
$h$, $H$ and $A$ is allowed at the energy scale of $m_A$ or $m_H$ and 
the new particles do not give significant contributions to 
Higgs-$\gamma\gamma$ vertices.

\subsection{Breit--Wigner Approximation}
\label{S4.1}

\begin{figure}[p]
\begin{center}
\hspace*{15mm}
\epsfxsize=7cm
\epsfysize=7cm
\epsffile{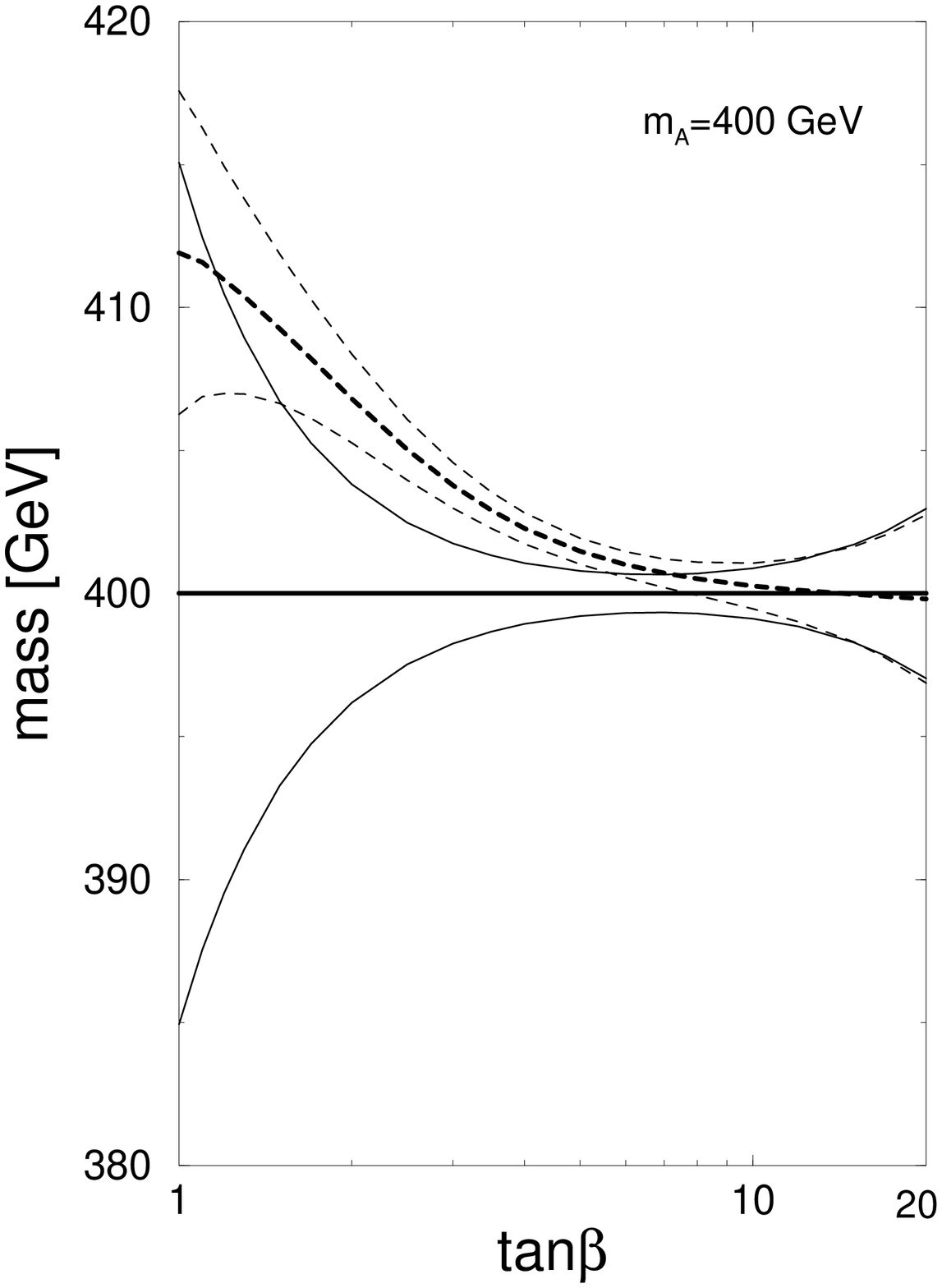}
\caption[Masses and widths of $H$ and $A$.]{\small 
 The masses of $H$ and $A$ are shown by bold-dashed curve and
bold-solid line, respectively, as functions of $\tan\beta$.  
 The total decay width of $H$ ($A$) is also indicated by the two 
thin-dashed (thin-solid) curves associated the mass curve, which 
draw $m_H \pm \Gamma_H$ ($m_A \pm \Gamma_A$).  
 Here $m_A$ is fixed to be 400 GeV.  }
\label{massfig}
\end{center}

\vfil

\begin{center}
\hspace*{15mm}
\epsfxsize=7cm
\epsfysize=7cm
\epsffile{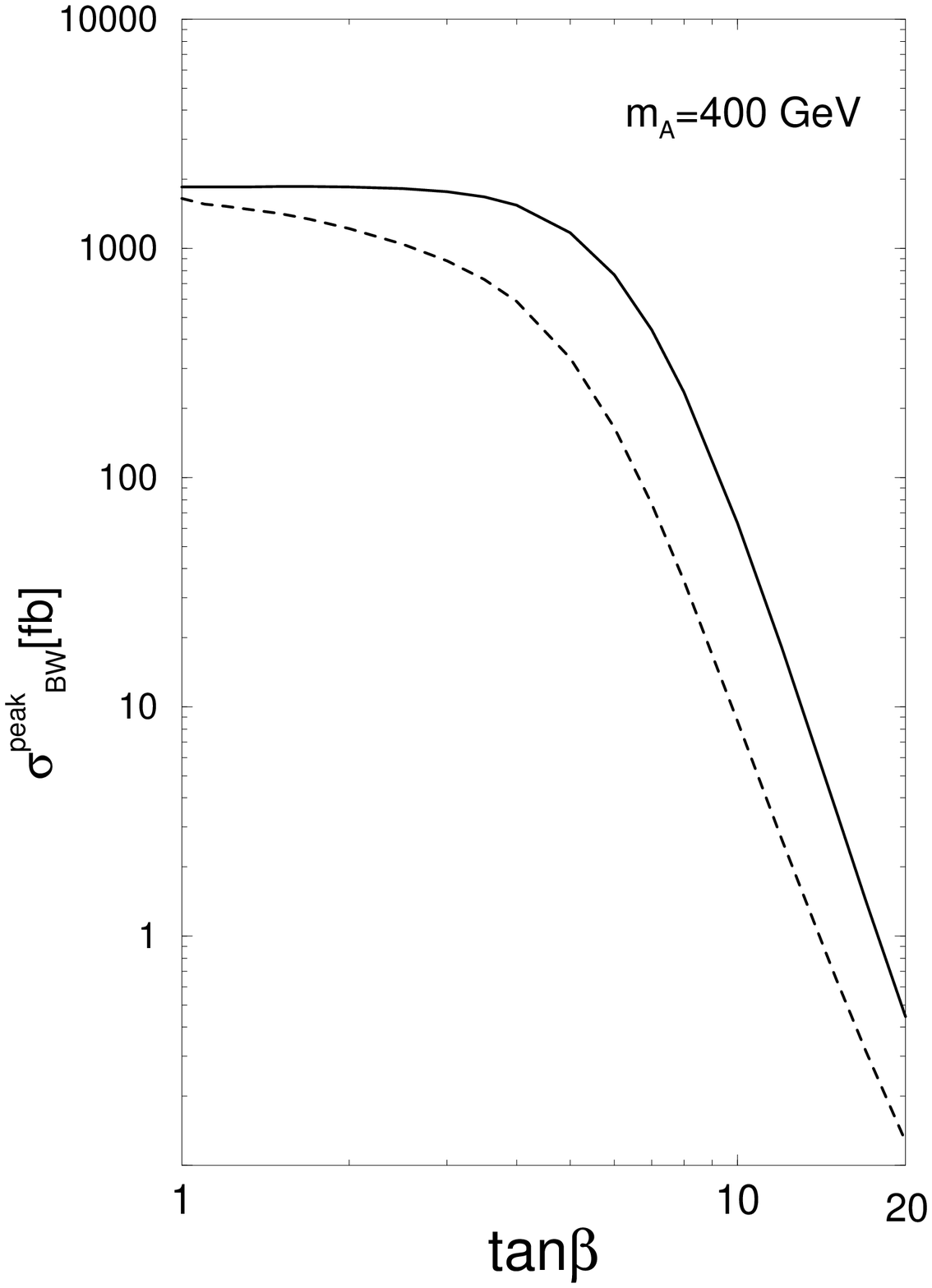}
\caption[Breit--Wigner peak cross sections.]{\small 
 The peak cross sections of the process $\gamma_+ \gamma_+ 
\rightarrow H,A \rightarrow t\overline{t}$ which are estimated by the 
Breit--Wigner approximation.  
 The dashed curve corresponds to $H$ and the solid curve to $A$.  
 Note that these computations neglect either of the continuum and 
the interference contributions.  }
\label{bwpeak}
\end{center}
\end{figure}

 A significant interference between the $H$-exchange and the 
$A$-exchange amplitudes can be expected with the small mass difference 
and the large decay widths of these Higgs bosons.  
 For $m_A$ = 400 GeV, Fig.~\ref{massfig} shows that the mass 
splitting between $H$ and $A$ is small for $\tan\beta$ $\sim$ 10.  
 The total decay widths of $H$ and $A$ ($\Gamma_H$ and $\Gamma_A$) 
are large for both $\tan\beta$ $\sim 1$ and $\tan\beta$ $\sim 10$, as 
are seen in Fig.~\ref{massfig}.  

 On the other hand, the peak cross sections of $\gamma\gamma 
\rightarrow H/A \rightarrow t\overline{t}$ computed by the 
Breit--Wigner approximation give a first guess in the magnitude of 
the Higgs contributions against the continuum contribution:  
\begin{equation}
\sigma_{\mbox{\tiny BW}}^{\mbox{\tiny peak}}
= 16\pi Br(H,A \rightarrow \gamma\gamma)\ 
        Br(H,A \rightarrow t\overline{t}) /m_{H,A}^{\;2} \ , 
\end{equation} 
where $Br$'s denote the decay branching ratios of $H$ or $A$ boson.  
 The Breit--Wigner peak cross sections of $H$ and $A$ are plotted in 
Fig.~\ref{bwpeak}.  
 As one can see from the figure, $\tan\beta$ $\simgt$ 10 results in 
suppression of these peak cross sections, which is due to the small 
branching ratios of both $H,A \rightarrow \gamma\gamma$ and 
$H,A \rightarrow t\overline{t}$.  
 We have chosen $\tan\beta$ = 3 and 7 in our simulation below. 
 The values of the masses, the total decay widths, the two-photon 
decay branching ratios and the $t\overline{t}$ branching ratios of 
$H$ and $A$ adopted in our further calculations are summarized in 
Table~\ref{masses}.

\subsection{Cross Sections of 
            \protect\boldmath $\gamma\gamma \rightarrow t\overline{t}$}
\label{S4.2}

\begin{figure}[p]
\hspace*{8mm}
\epsfxsize=150mm
\epsfysize=150mm
\epsffile{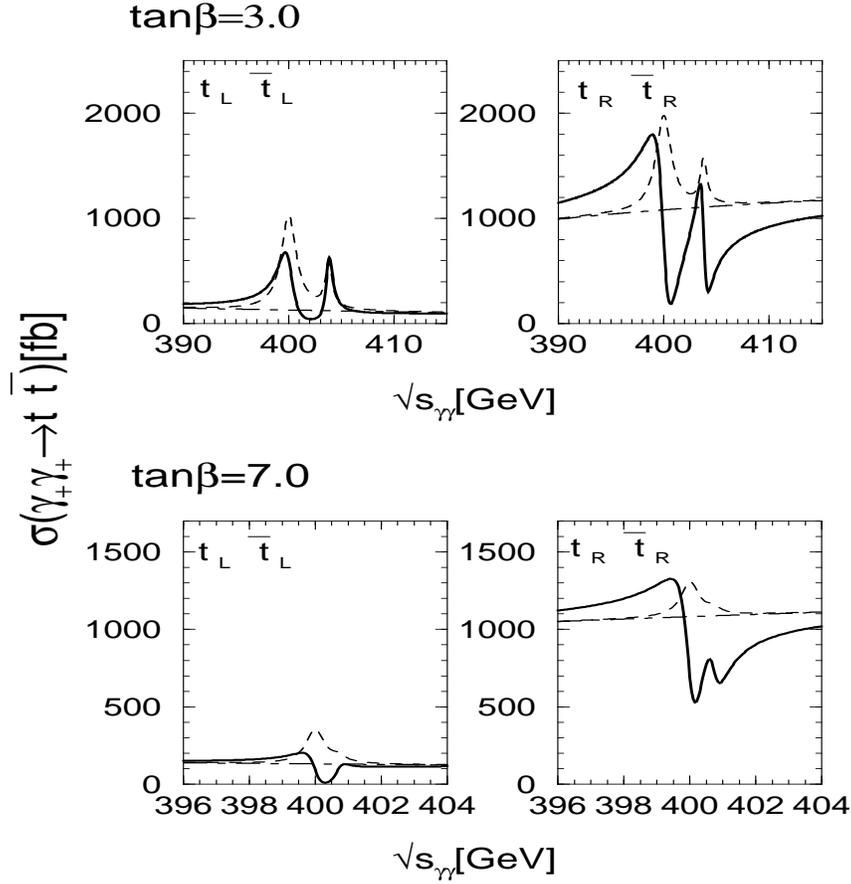}
\vspace*{-15mm}
\caption[Cross sections]{\small 
 The cross sections of $\gamma_+ \gamma_+ \rightarrow 
t_L \overline{t}_L$ (left), $t_R \overline{t}_R$ (right) for 
$\tan\beta$ = 3 (above) and 7 (below).  
 The solid curves show the cross sections with Higgs resonances 
$\sigma^{++ \lambda \overline{\lambda}}$, the dashed curves the 
ones obtained by neglecting the interference 
$\sigma_0^{++ \lambda \overline{\lambda}}$, 
the dot-dashed lines the continuum contributions 
$\sigma_{\mbox{\tiny tree}}^{++ \lambda \overline{\lambda}}$, 
respectively.  }
\label{cross}
\end{figure}

 Fig.~\ref{cross} shows the energy dependence of $\gamma_+ \gamma_+ 
\rightarrow t_{\lambda} \overline{t}_{\overline{\lambda}}$ cross 
section for $\tan\beta$ = 3 and 7.  
 The solid curves represent the cross sections with the interference; 
$\sigma^{++ \lambda \overline{\lambda}}$, see eq.~(\ref{10a}), while 
the dashed curves show the one obtained by neglecting the 
interference; $\sigma_0^{++ \lambda \overline{\lambda}}$, see 
eq.~(\ref{10b}).  
 The cross sections without any Higgs production are also
superimposed as in the dot-dashed curves; 
$\sigma^{++\lambda \overline{\lambda}}_{\mbox{\tiny tree}}$,
see eq.~(\ref{10c}).  
 The manifest definitions of these cross sections are as follows:  
\begin{subequations}
\begin{eqnarray}
\sigma^{++ \lambda \overline{\lambda}} 
 & = & 
\frac{N_c \beta_t}{32\pi s_{\gamma\gamma}} 
\int_{-1}^{+1} d\/\cos\theta_t \, 
\bigl| {\cal M}_{\gamma\gamma \rightarrow t\overline{t}}%
 ^{++ \lambda \overline{\lambda}} \bigr|^2 \ , 
\label{10a} \\
\sigma_0^{++ \lambda \overline{\lambda}} 
 & = & 
\frac{N_c \beta_t}{16\pi s_{\gamma\gamma}} \biggl[
\bigl| {\cal M}_H^{++ \lambda \overline{\lambda}} \bigr|^2 + 
\bigl| {\cal M}_A^{++ \lambda \overline{\lambda}} \bigr|^2 \biggr] 
+ \sigma_{\mbox{\tiny tree}}^{++ \lambda \overline{\lambda}} \ , 
\label{10b} \\
\sigma_{\mbox{\tiny tree}}^{++ \lambda \overline{\lambda}} 
 & = & 
\frac{N_c \beta_t}{32\pi s_{\gamma\gamma}} 
\int_{-1}^{+1} d\/\cos\theta_t \, 
\bigl| {\cal M}_{\mbox{\tiny tree}}^{++ \lambda \overline{\lambda}} 
\bigr|^2 \ , 
\label{10c}
\end{eqnarray}
\label{defcross}
\end{subequations}
\hspace*{-0.7em}
where $N_c$ is the color factor of the top quark and 
$s_{\gamma\gamma}$ is the collision energy squared.  
 As one can see in the figures, the interference is constructive at 
the energy below $m_A$, while it is destructive above $m_A$.  
 Especially, the cross sections for the $(++RR)$ case, $\sigma^{++RR}$, 
is strongly variant and have sharp troughs, which show us drastic 
effects of the interference.  

\begin{figure}[t]
\begin{center}
\hspace*{10mm}
\epsfxsize=13cm
\epsfysize=10cm
\epsffile{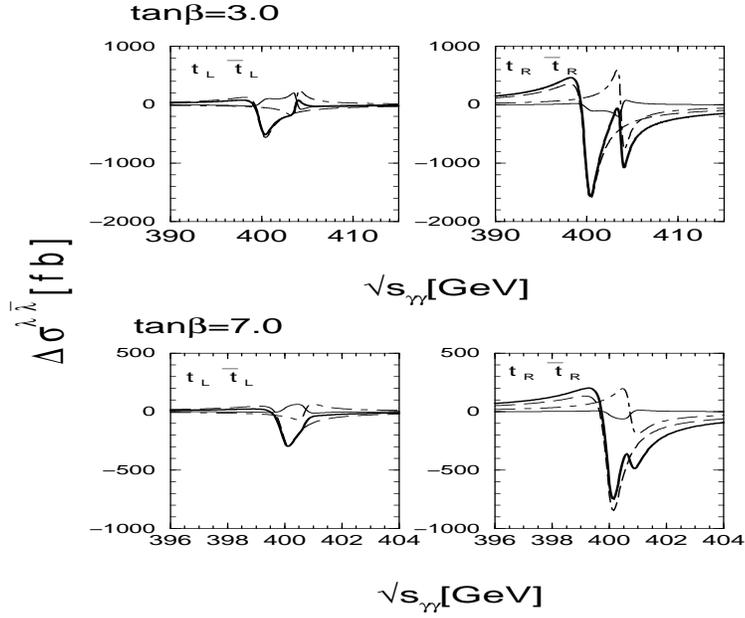}
\caption[Interferences]{\small 
 The interference contributions of the cross sections of 
$\gamma_+ \gamma_+ \rightarrow t_L \overline{t}_L$ (left), 
$t_R \overline{t}_R$ (right) for $\tan\beta$ = 3 (above) and 7 
(below).  
 The thin-solid, dot-dashed and dashed curves show 
$\Delta\sigma_{\mbox{\tiny $A$-$H$}}%
^{++ \lambda \overline{\lambda}}$, 
$\Delta\sigma_{\mbox{\tiny $H$-tree}}%
^{++ \lambda \overline{\lambda}}$ 
and 
$\Delta\sigma_{\mbox{\tiny $A$-tree}}%
^{++ \lambda \overline{\lambda}}$, 
respectively, and the bold-solid ones mean the sum of these three, 
$\Delta\sigma^{++ \lambda \overline{\lambda}}$.  }
\label{int}
\end{center}
\end{figure}

 As described above, the interference term consists of three distinct 
contributions; {\it i.e.}, $A$-$H$, $H$-tree and $A$-tree.  
 We define the following four quantities representing the interference 
contributions:  
\begin{subequations}
\begin{eqnarray}
\Delta\sigma^{++ \lambda \overline{\lambda}}
& = & \sigma^{++ \lambda \overline{\lambda}} 
    - \sigma_0^{++ \lambda \overline{\lambda}} 
\ = \ 
  \Delta\sigma_{\mbox{\tiny $A$-$H$}}^{++ \lambda \overline{\lambda}}
+ \Delta\sigma_{\mbox{\tiny $H$-tree}}^{++ \lambda \overline{\lambda}} 
+ \Delta\sigma_{\mbox{\tiny $A$-tree}}^{++ \lambda \overline{\lambda}} 
\ , 
\label{14a} \\
\Delta\sigma_{\mbox{\tiny $A$-$H$}}^{++ \lambda \overline{\lambda}}
& = & \frac{N_c \beta_t}{8 \pi s_{\gamma\gamma}} \ {\Re}e 
\big({\cal M}_A^{++ \lambda \overline{\lambda}} \ 
     {\cal M}_H^{++ \lambda \overline{\lambda}} {}^* 
\big) \ , 
\label{14b} \\
\Delta\sigma_{\mbox{\tiny $H$-tree}}^{++ \lambda \overline{\lambda}} 
& = & \frac{N_c \beta_t}{16 \pi s_{\gamma\gamma}} 
\int_{-1}^{+1} d\/\cos\theta_t \ {\Re}e 
\big({\cal M}_H^{++ \lambda \overline{\lambda}} \ 
     {\cal M}_{\mbox{\tiny tree}}^{++ \lambda \overline{\lambda}} {}^* 
\big) \ , 
\label{14d} \\
\Delta\sigma_{\mbox{\tiny $A$-tree}}^{++ \lambda \overline{\lambda}} 
& = & \frac{N_c \beta_t}{16 \pi s_{\gamma\gamma}} 
\int_{-1}^{+1} d\/\cos\theta_t \ {\Re}e 
\big({\cal M}_A^{++ \lambda \overline{\lambda}} \ 
     {\cal M}_{\mbox{\tiny tree}}^{++ \lambda \overline{\lambda}} {}^* 
\big) \ .  
\label{14c}
\end{eqnarray}
\label{diff}
\end{subequations}
\hspace*{-0.8em}
 These four values are plotted in Fig.~\ref{int}.  
 The $A$-tree term gives the dominant interference effect in the 
(++LL) channel, whereas both the $A$-tree and $H$-tree terms are 
significant in the (++RR) channel.  

 The sign of the dominant Higgs-tree components 
$\Delta\sigma_{\mbox{\tiny $A$-tree}}^{++ \lambda \overline{\lambda}}$ 
and 
$\Delta\sigma_{\mbox{\tiny $H$-tree}}^{++ \lambda \overline{\lambda}}$ 
is dependent on the complex phases of the 
$\sum_i I^i_{H,A}$ factor in eqs.~(\ref{loop-h}), (\ref{loop-a}) and 
of the propagator ${\cal B}_{H,A}$ in eq.~(\ref{circle}).  
 The $\sqrt{s}_{\gamma\gamma}$ dependence of the signs of the 
interference contributions from the Higgs-tree components are 
summarized in Table~\ref{tab:exp}, for the cases adopted in our 
numerical simulation, where $-\pi/2 < \arg \sum_i I^i_H < \pi/2$ and 
$\pi/2 < \arg \sum_i I^i_A < 3\pi/2$.  
 From this table, one can qualitatively understand the energy 
dependence of $\Delta\sigma^{++ \lambda \overline{\lambda}}$ found in 
Fig.~\ref{int}.

\subsection{Convoluted Cross Sections with 
            \protect\boldmath $\gamma\gamma$ Luminosity}
\label{S4.3}

\begin{figure}[b]
\begin{center}
\hspace*{0mm}
\epsfxsize=7cm
\epsffile{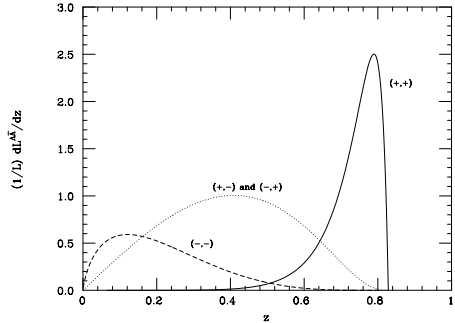}
\caption[Photon-photon luminosity]{\small 
 The luminosity distributions of the $\gamma\gamma$ colliders.  
 The solid, dotted and dashed curves are correspond to the 
$\gamma\gamma$ collisions with the helicities of $++$, $+-$ and $-+$, 
and $--$, respectively.  
 The horizontal axis $z$ means the energy fraction of the colliding 
photons $E_{\gamma\gamma}$ to the original electron collider 
$\sqrt{s}_{ee}$.  }
\label{lumi}
\end{center}
\end{figure}

 Due to the spread collision energy of $\gamma\gamma$ colliders, the 
observed cross section is a convoluted one with the $\gamma\gamma$ 
luminosity.  
 A detailed study on the possible luminosity and polarization 
distributions at future $\gamma\gamma$ colliders has been performed by 
the simulation program CAIN[\ref{CAIN}].  
 However, these parameters are dependent on the machine design of the 
colliders.  
 Thus we have adopted an ideal situation of the beam conversion that 
the photon beam is generated by the tree-level formula of the Compton 
backward-scattering and the effect of the finite scattering angle is 
negligible [\ref{compton}], 
\begin{subequations}
\begin{eqnarray}
D(y) & = & \frac{D_1 + P_e P_L D_2}{D_3 + P_e P_L D_4} \ , \\
P(y) & = & \frac{P_e P_1 + P_L P_2}{D_1 + P_e P_L D_2} \ , \\
D_1 & = & 1-y+1/(1-y)-4r(1-r) \ , \\
D_2 & = & -rx(2-y)(2r-1) \ , \\
D_3 & = & (1-4/x-8/x^2)\log(x+1)+1/2+8/x^2-1/2(x+1)^2 \ , \\
D_4 & = & (1+2/x)\log(x+1)-5/2+1/(x+1)-1/2(x+1)^2 \ , \\
P_1 & = & rx(1+(1-y)(2r-1)^2 \ , \\
P_2 & = & -(1-y+1/(1-y))(2r-1) \ , 
\end{eqnarray}
\label{DP}
\end{subequations}
\hspace*{-0.9em}
where $D(y)$ and $P(y)$ are the energy and the circular polarization 
distributions, $y$ the energy fraction of the obtained beam photon to 
the original beam electron, $x$ the squared ratio of the total energy 
of the Compton scattering in the center-of-mass system to the 
electron mass, $r$ = $y/x(1-y)$, and $P_e$ and $P_L$ the 
polarizations of the original electron beam and the laser photon beam, 
respectively.  
 Here we have adopted the optimum value of $x$, $2+2\sqrt{2}$ 
[\ref{ggoriginal}], and assumed ideal polarizations of the original 
beams, {\it i.e.}, $P_e$ = $+1$ and $P_L$ = $-1$ for both sides.  
 The luminosity distribution of the colliding $\gamma\gamma$ in 
each photon helicities can be evaluated as follows:  
\begin{equation}
\frac{1}{\cal L} \frac{d {\cal L}^{\Lambda \overline{\Lambda}}}{dz} 
= \int_{\log z/z_{\mbox{\tiny max}}}^{\log z_{\mbox{\tiny max}}} 
d\eta \ 2z 
D(z e^{\eta}) 
\frac{1 + \Lambda P(z e^{\eta})}{2} \ 
D(z e^{-\eta}) 
\frac{1 + \overline{\Lambda} P(z e^{-\eta})}{2} \ , 
\label{conv}
\end{equation}
where $z = \sqrt{s}_{\gamma\gamma}/\sqrt{s}_{ee}$, 
$z_{\mbox{\tiny max}}$ $= x/(x+1)$ $= 2\sqrt{2}-2$ is the maximum 
value of $z$, $\eta$ is the rapidity of the colliding $\gamma\gamma$ 
system in the laboratory frame, $\cal L$ is the total $\gamma\gamma$ 
luminosity%
\footnote{
 In this approximation, the right-hand-side of eq.~(\ref{conv}) 
integrated over $z$ from 0 to $z_{\mbox{\tiny max}}$ and summed up 
all $\Lambda \overline{\Lambda}$ combinations is unity.  
}.  
 Resulting $\gamma\gamma$ luminosities are illustrated in 
Fig.~\ref{lumi} separately for the combination of the beam photon 
helicities.  

\begin{figure}[t]
\begin{center}
\hspace*{-5mm}
\epsfxsize=135mm
\epsfysize=100mm
\epsffile{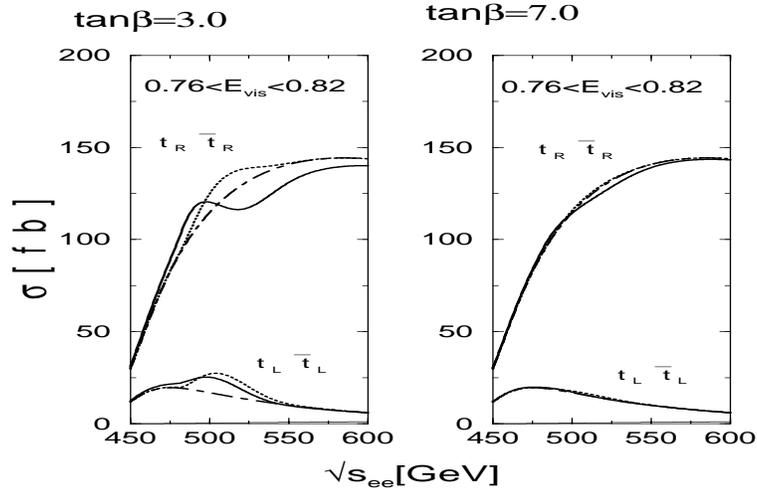}
\caption[Luminosity convoluted cross sections]{\small 
 The effective cross sections convoluted by the $\gamma\gamma$ 
luminosity with the visible energy cut are illustrated.  
 The bold-solid curves correspond to the correct cross sections 
$\sigma^{\lambda \overline{\lambda}}$, while dotted and dot-dashed 
ones to $\sigma_0^{\lambda \overline{\lambda}}$ and 
$\sigma_{\mbox{\tiny tree}}^{\lambda \overline{\lambda}}$, 
respectively.  
 The upper curves are for $t_R \overline{t}_R$, and the lower ones for 
$t_L \overline{t}_L$.  
 The sum of the tree cross sections for $t_R \overline{t}_L$ and 
$t_L \overline{t}_R$, $\sigma^{RL+LR}$, are also plotted in the 
thin-solid line located very near to the bottom horizontal axis.  
 The left figure is for $\tan\beta$ = 3, and the right for $\tan\beta$ 
= 7.  }
\label{lumicross}
\end{center}
\end{figure}

 One can evaluate the observed cross section by convoluting the cross 
section of the bare process calculated in the previous subsection with 
the above luminosity distribution, and by taking account of smearing 
due to the resolution of the detector.  
 The resulting `effective' cross section 
$\sigma^{\lambda \overline{\lambda}}$ is expressed as, 
\begin{eqnarray}
\sigma^{\lambda \overline{\lambda}} 
(\sqrt{s}_{ee};\ E_n, E_m) & = & 
\int_{E_n}^{E_m} d E_{\mbox{\tiny vis}} \  
\int_{2m_t}^{z_{\mbox{\tiny max}} 
\sqrt{s}_{ee}} d\sqrt{s}_{\gamma\gamma} 
\sum_{\Lambda \overline{\Lambda}} 
\frac{1}{\cal L} \frac{d{\cal L}^{\Lambda \overline{\Lambda}}}%
 {d\sqrt{s}_{\gamma\gamma}} 
\nonumber \\ 
 & & \quad \times 
\hat{\sigma}^{\Lambda \overline{\Lambda} \lambda \overline{\lambda}}
 (\sqrt{s}_{\gamma\gamma}) \ 
G(\sqrt{s}_{\gamma\gamma} - E_{\mbox{\tiny vis}},r) \, 
\label{smear}
\end{eqnarray}
where $E_{\mbox{\tiny vis}}$ is the visible energy in the detector, 
$E_n$ and $E_m$ the minimum and the maximum visible energy cuts, 
$r$ is the detector resolution on measuring 
$\sqrt{s}_{\gamma\gamma}$ and $G$ is the probability distribution of 
the detector resolution smearing.  
 Here after, the bare cross section 
$\sigma^{\Lambda \overline{\Lambda} \lambda \overline{\lambda}}$ 
in the previous subsection is denoted as
$\hat{\sigma}^{\Lambda \overline{\Lambda} \lambda \overline{\lambda}}$, 
as in the right-hand-side of eq.~(\ref{smear}).  
 We have assumed the Gaussian distribution $G(x,y)$ 
$= \exp(-x^2/2y^2)/\sqrt{2\pi}y$, with the detector resolution 
$r/\sqrt{s}_{\gamma\gamma}$ 
$= \sqrt{40{\%}^2/(\sqrt{s}_{\gamma\gamma}/{\mbox{GeV}})+2{\%}^2}$, 
which is the one for the standard hadron calorimeter in the JLC 
design [\ref{JLC}].
 In Fig.~\ref{lumicross}, the effective cross sections for various 
combinations of the $t\overline{t}$ helicities are drawn as functions 
of $\sqrt{s}_{ee}$.  
 The cut energies adopted here are $E_n$ = $0.76 \sqrt{s}_{ee}$ and 
$E_m$ = $0.82 \sqrt{s}_{ee}$ in order to use the energy range of the 
differential luminosity, $d{\cal L}/dz$, where both colliding photon 
beams are well polarized to be almost $+100${\%}.  
 For the case of $\tan\beta$ = 3, one can clearly observe the 
existence of the Higgs bosons at both $LL$ and $RR$ modes, in 
comparison with the continuum cross sections.  
 The effects of the interference terms are also sizable even in the 
convoluted effective cross sections at $\tan\beta$ = 3.  
 These effects are destructive at $\sqrt{s}_{ee}$ $<$ 498 GeV for 
$\sigma^{RR}$ and $<$ 491 GeV for $\sigma^{LL}$, and are constructive 
above these collider energies.  
 Especially, the constructive contribution of the interference terms 
in $\sigma^{RR}$ is remarkably large at the wide range of the 
$\sqrt{s}_{ee}$ above $\sim$500 GeV.  
 Thus there is a possibility that one can identify two Higgs signals 
separately by a careful analysis scanning $\sqrt{s}_{ee}$.  
 In the case of $\tan\beta$ = 7, unfortunately, the signals of $H$ 
and $A$ are almost smeared out, since the magnitudes of the 
Higgs-exchanging amplitudes are small and the mass gap of $A$ and $H$ 
is narrow.  
 For both values of $\tan\beta$, contributions coming from continuum 
$t_L \overline{t}_R$ and $t_R \overline{t}_L$ are negligibly small 
due to the suppressed ${\cal L}^{+-}$ and ${\cal L}^{-+}$ 
luminosities between the visible energy cuts, $E_n$ and $E_m$.

\subsection{CP-parity Measurement of the Higgs Bosons}
\label{S4.4}

 One can extract information about the CP-parity of Higgs bosons by 
observing the interference effects on the cross sections with the 
top helicity fixed%
\footnote{
The statistical tachnique to measure the top helicity is 
discussed in the sub-subsection~\ref{S4.5.2}.  }, 
$\sigma^{RR}$ and $\sigma^{LL}$.  
 Here we present a convenient method to estimate the CP-parity of 
the Higgs boson from the measurements of $\sigma^{RR}$ and 
$\sigma^{LL}$.  

 First, for simplicity, let us assume only one Higgs boson of $H$ or 
$A$.  
 The cross section of the bare process $\gamma_+ \gamma_+ \rightarrow 
t_\lambda \overline{t}_{\overline{\lambda}}$ with a Higgs resonance, 
$\hat{\sigma}^{++\lambda\overline{\lambda}}$, deviates from the 
continuum one, 
$\hat{\sigma}^{++\lambda\overline{\lambda}}_{\mbox{\tiny tree}}$.  
 We can define the deviation as, 
\begin{subequations}
\begin{eqnarray}
\Delta \hat{\sigma}_t^{++\lambda\overline{\lambda}} 
 & \equiv & 
\hat{\sigma}^{++ \lambda \overline{\lambda}}
- \hat{\sigma}_{\mbox{\tiny tree}}^{++ \lambda \overline{\lambda}} 
\ = \ \Delta \hat{\sigma}_{\mbox{\tiny Higgs}} 
+ \Delta \hat{\sigma}_{\mbox{\tiny Higgs-tree}}%
^{++\lambda\overline{\lambda}} 
\ , \label{excess} \\
\Delta \hat{\sigma}_{\mbox{\tiny Higgs}} 
 & = & 
\frac{N_c \beta_t}{16 \pi s_{\gamma\gamma}} 
\bigl| {\cal M}_{\mbox{\tiny Higgs}}^{++ \lambda \overline{\lambda}} 
\bigr|^2 \ , 
\label{excess-h} \\
\Delta \hat{\sigma}_{\mbox{\tiny Higgs-tree}}%
^{++\lambda\overline{\lambda}} 
 & = & 
\frac{N_c \beta_t}{16 \pi s_{\gamma\gamma}} 
{\Re}e \big( 
{\cal M}_{\mbox{\tiny Higgs}}^{++ \lambda \overline{\lambda}} \big) 
\int_{-1}^{+1} d\/\cos\theta_t \, 
{\cal M}_{\mbox{\tiny tree}}^{++ \lambda \overline{\lambda}} \ , 
\label{excess-i}
\end{eqnarray}
\end{subequations}
where eqs.~(\ref{excess-h}) and (\ref{excess-i}) are the 
contributions of the Higgs-exchange amplitude squared and the 
interference.  
 According to the discussion on the CP-parity of the Higgs bosons in 
Secs.~\ref{S2} and \ref{S3}, one can see that 
$\Delta \hat{\sigma}_{\mbox{\tiny Higgs-tree}}^{++RR}$ and 
$\Delta \hat{\sigma}_{\mbox{\tiny Higgs-tree}}^{++LL}$ have the same 
signs for a pseudoscalar $A$, while 
$\Delta \hat{\sigma}_{\mbox{\tiny Higgs-tree}}^{++RR}$ and 
$\Delta \hat{\sigma}_{\mbox{\tiny Higgs-tree}}^{++LL}$ have the 
opposite signs for a scalar $H$.  
 Since the tree amplitudes satisfy an inequality, 
\begin{equation}
\bigl| {\cal M}_{\mbox{\tiny tree}}^{++LL} \bigr| < 
\bigl| {\cal M}_{\mbox{\tiny tree}}^{++RR} \bigr| \ , 
\end{equation}
then, one can derive, 
\begin{equation}
\bigl| \Delta \hat{\sigma}_{\mbox{\tiny Higgs-tree}}^{++LL} \bigr| < 
\bigl| \Delta \hat{\sigma}_{\mbox{\tiny Higgs-tree}}^{++RR} \bigr| 
\ .  
\end{equation}
 Since the interference term 
$\Delta \hat{\sigma}_{\mbox{\tiny Higgs-tree}}%
^{++\lambda \overline{\lambda}}$ 
is suppressed more mildly than the Higgs term 
$\Delta \hat{\sigma}_{\mbox{\tiny Higgs}}$ far away from the 
resonance, 
$\Delta \hat{\sigma}_t^{++\lambda\overline{\lambda}}$ is negative at 
one side of $\sqrt{s}_{\gamma\gamma} \ll m_{\mbox{\tiny Higgs}}$ and 
$\sqrt{s}_{\gamma\gamma} \gg m_{\mbox{\tiny Higgs}}$.  
 At the $\sqrt{s}_{\gamma\gamma}$ where $\Delta \hat{\sigma}_t^{++LL}$ 
is negative, we can conclude the following inequalities of judgment:  
\begin{equation}
\Delta \hat{\sigma}_t^{++LL} < 0 \quad \Rightarrow \quad \left\{ 
\begin{array}{ll}
\Delta \hat{\sigma}_t^{++RR} > 0 & \mbox{for $H$}, \\
\Delta \hat{\sigma}_t^{++RR} < 0 & \mbox{for $A$}.  
\end{array}
\right. 
\label{judge}
\end{equation}
Therefore, we can distinguish the CP-parity of the Higgs boson.  

 Unfortunately, it is not trivial whether this CP-parity judgment 
method always works after convoluting with the $\gamma\gamma$ 
luminosity, because the broad luminosity distribution may smear out 
information about the sign of the interference term.  
 If there are plural Higgs bosons with near values of masses, as are 
often happen in the MSSM, the judgment may be obscured.  
 Actually, in our mumerical simulation in the previous 
subsection~\ref{S4.3}, $\Delta \hat{\sigma}_t^{++LL}$ is negative 
only at $\sqrt{s}_{ee}$ $>$ 545 and 514 GeV for the $\tan\beta$ = 3 
and 7 cases, respectively.  
 And at the same energy range, $\Delta \hat{\sigma}_t^{++RR}$ is 
also negative for both cases, which means the existence of a 
pseudoscalar Higgs $A$.  
 In these cases, the effect of the scalar Higgs $H$ is smaller than 
that of $A$, and one cannot find out the signal of $H$ existence 
by this method.

\subsection{Number Count Estimates}
\label{S4.5}

\subsubsection{Higgs boson detection}

 The existence of Higgs bosons can be verified by observing the 
excess of the total cross section due to the Higgs resonances 
over the continuum.  
 The estimated number of events of the process $\gamma\gamma 
\rightarrow t\overline{t}$, $N(t\overline{t})$, is proportional to 
the integrated luminosity $\int \! dt {\cal L}$ and the detection 
efficiency $\epsilon$, 
\begin{equation}
N(t\overline{t}) = \sigma \cdot \epsilon \int \!\! dt {\cal L} \ , 
\end{equation}
where, 
\begin{subequations}
\begin{eqnarray}
\sigma \qquad \ \ 
 & = & \sigma^{RR} + \sigma^{LL} + \sigma^{RL+LR} \ , \\
\sigma^{RL+LR} & = & \sigma^{RL} + \sigma^{LR} \ .  
\end{eqnarray}
\label{star}
\end{subequations}

 The statistical error of the cross section $E(\sigma)$ is 
estimated to be 
$\sqrt{N(t\overline{t})}/\epsilon \int \! dt {\cal L}$, and hence 
the relative error of the total cross section, ${\cal E}(\sigma)$, 
is inversely proportional to $\sqrt{\epsilon \int \! dt {\cal L}}$:  
\begin{equation}
{\cal E}(\sigma) \equiv \frac{E(\sigma)}{\sigma} 
 = \frac{1}{\sqrt{N(t\overline{t})}} 
 = \frac{1}{\sqrt{\sigma \cdot \epsilon \int \! dt {\cal L}}} \ .  
\label{esigma}
\end{equation}
 Since the detection efficiency $\epsilon$ is not studied well for 
$\gamma\gamma$ colliders, we do not assume its particular numerical 
value.  
 Instead, we define $\hat{\cal E}(\sigma)$ which is the special value 
of ${\cal E}(\sigma)$ with $\epsilon \int \! dt {\cal L}$ = 1 
fb$^{-1}$.  
 The values of the total cross section $\sigma$ and its statistical 
error for the nominal integrated luminosity $\hat{\cal E}(\sigma)$ 
are listed in 
Table~\ref{total} for nine particular values between $\sqrt{s}_{ee}$ 
= 480 and 560 GeV.  
 The statistical error ${\cal E}(\sigma)$ is smaller than 
10{\%} for all the cases listed in Table~\ref{total}, if we assume 1 
fb$^{-1}$ of $\epsilon \int \! dt {\cal L}$.  

 The existence of the Higgs signals can be studied by comparing the 
measured effective cross section $\sigma$ with the computed one 
without Higgs resonances, $\sigma_{\mbox{\tiny tree}}$, which is 
defined similarly to $\sigma$.  
 The statistical significance of the deviation of 
$\sigma$ from $\sigma_{\mbox{\tiny tree}}$ can be evaluated as, 
\begin{equation}
{\cal S} \equiv 
\frac{\sigma - \sigma_{\mbox{\tiny tree}}}%
{E(\sigma)} 
= \frac{\sigma - \sigma_{\mbox{\tiny tree}}}{\sqrt{\sigma}} \ 
\sqrt{\epsilon \int \!\! dt {\cal L}} \ , 
\label{signif}
\end{equation}
where the sign of ${\cal S}$ indicates the sign of deviation.  
 Since ${\cal S}$ is proportional to $\sqrt{\epsilon \int \! dt 
{\cal L}}$, we define $\hat{\cal S}$ which is the special value of 
${\cal S}$ with 1 fb$^{-1}$ of $\epsilon \int \! dt {\cal L}$.  
 The values of $\sigma_{\mbox{\tiny tree}}$ and $\hat{\cal S}$ 
are also listed in Table~\ref{total}.  
 As is seen in the table, the effective cross section $\sigma$ 
deviates from $\sigma_{\mbox{\tiny tree}}$ in the one-sigma level at 
$\sqrt{s}_{ee}$ = 480---500 and 530---540 GeV for $\tan\beta$ = 3, 
even with 1 fb$^{-1}$ of $\epsilon \int \! dt {\cal L}$.  
 Tuning $\sqrt{s}_{ee}$ = 490 GeV, one can establish the existence of 
the Higgs resonances at the two-sigma level with 1.6 fb$^{-1}$.  
 On the other hand, we need more than 40 fb$^{-1}$ at $\sqrt{s}_{ee}$ 
= 520 GeV in the case of $\tan\beta$ = 7.

\subsubsection{CP-parity measurement}
\label{S4.5.2}

 In order to observe the CP-parity of the Higgs boson by the method 
described in the subsection~\ref{S4.4}, we need to measure 
$\sigma^{RR}$ and $\sigma^{LL}$ separately.  
 The measurement of the top helicity can be performed statistically 
by using the decay angle dependence of the polarized top quark.  
 In the tree-level approximation with massless bottom quark, one can 
easily derive the decay angle dependence for each of $t_R$ and 
$t_L$:  
\begin{subequations}
\begin{eqnarray}
\frac{1}{\Gamma_{t_R}} \frac{d \, \Gamma_{t_R}}{d\cos\theta_b} = 
\frac{1}{\Gamma_{\overline{t}_L}} \frac{d \, \Gamma_{\overline{t}_L}}%
{d\cos\theta_b} & = & 
 \frac{1}{2} 
 \left[ 1 - \frac{m_t^{\;2} - 2 m_W^{\;2}}{m_t^{\;2} + 2 m_W^{\;2}} 
        \cos\theta_b \right] 
 \simeq 0.50 - 0.20 \cos\theta_b \ , \\
\frac{1}{\Gamma_{t_L}} \frac{d \, \Gamma_{t_L}}{d\cos\theta_b} = 
\frac{1}{\Gamma_{\overline{t}_R}} \frac{d \, \Gamma_{\overline{t}_R}}%
{d\cos\theta_b} & = & 
 \frac{1}{2} 
 \left[ 1 + \frac{m_t^{\;2} - 2 m_W^{\;2}}{m_t^{\;2} + 2 m_W^{\;2}} 
        \cos\theta_b \right] 
 \simeq 0.50 + 0.20 \cos\theta_b \ , 
\end{eqnarray}
\label{decay}
\end{subequations}
\hspace*{-0.9em}
where $\cos\theta_b$ is the emission angle of the (anti-)bottom 
quark in the rest frame of the decaying (anti-)top quark with respect 
to the direction of the (anti-)top momentum in the $t\overline{t}$ 
c.m.\ frame.  
 This decay angle dependence of the top quark allows us a simple 
method to estimate the top helicity, without studying the subsequent 
$W$ boson decay topology.  
 The fraction of the forward ($0$ $< \cos\theta_b$ $< 1$) decay of 
the right-handed top quark $r_F$ is 0.40, while the backward ($-1$ 
$< \cos\theta_b$ $< 0$) fraction $r_B$ is 0.60, where we neglect the 
effects of the higher order diagrams and the finite resolution.  

 An event with the top quark decaying forward and with the anti-top 
quark decaying backward is most likely to be a $t_L \overline{t}_L$ 
event because of the above decay distributions.  
 We define here two effective cross sections with decay angle 
selections:  {\it i.e.} $\sigma_F$ with the forward top and the 
backward anti-top, and $\sigma_B$ with the backward top and the 
forward anti-top.  
 By neglecting the miss-identification probability, we obtain, 
\begin{subequations}
\begin{eqnarray}
\sigma_F & = & 
 \sigma^{RR} r_F^{\;2} + \sigma^{LL} r_B^{\;2} 
 + \sigma^{RL+LR} r_F r_B \ , \\
\sigma_B & = & 
 \sigma^{RR} r_B^{\;2} + \sigma^{LL} r_F^{\;2} 
 + \sigma^{RL+LR} r_F r_B \ .  
\end{eqnarray}
\end{subequations}
 In Table~\ref{fb}, $\sigma^{RR}$, $\sigma^{LL}$, $\sigma^{RL+LR}$, 
$\sigma_F$ and $\sigma_B$ are listed for nine values of 
$\sqrt{s}_{ee}$ and at $\tan\beta$ = 3 and 7.  
 Since $\sigma^{RL+LR}$ is small enough, we can neglect this 
contribution and can evaluate the values of the intrinsic effective 
cross sections $\sigma^{RR}$ and $\sigma^{LL}$ from $\sigma_F$ and 
$\sigma_B$ as, 
\begin{equation}
\left( 
\begin{array}{@{\,}c@{\,}}
\sigma^{RR} \\
\sigma^{LL} 
\end{array}
\right) 
 \simeq 
\frac{1}{r_B^{\;4} - r_F^{\;4}} 
\left(
\begin{array}{@{\,}cc@{\,}}
-r_F^{\;2} &  r_B^{\;2} \\
 r_B^{\;2} & -r_F^{\;2} 
\end{array}
\right) 
\left( 
\begin{array}{@{\,}c@{\,}}
\sigma_F \\
\sigma_B 
\end{array}
\right) \ .  
\label{inverse}
\end{equation}
 The validity of neglecting $\sigma^{RL+LR}$ is estimated, and the  
shift to the measured cross section by this approximation from the 
true value is found to be less than 0.4 fb ($\sim$ 4.2{\%}) for the 
cases listed in Table~\ref{fb}.  

 The relative errors of the $\sigma^{RR}$ and $\sigma^{LL}$, 
${\cal E}(\sigma^{RR})$ and ${\cal E}(\sigma^{LL})$, are 
similarly defined as ${\cal E}({\sigma})$ in eq.~(\ref{esigma}), 
and can be evaluated as, 
\begin{subequations}
\begin{eqnarray}
{\cal E}(\sigma^{RR}) & = & 
 \frac{\sqrt{\sigma_F r_F^{\;4} + \sigma_B r_B^{\;4}}}%
      {\left| \sigma_F r_F^{\;2} - \sigma_B r_B^{\;2} \right|} 
 \frac{1}{\sqrt{\epsilon' \int \! dt {\cal L}}} \ , \\
{\cal E}(\sigma^{LL}) & = & 
 \frac{\sqrt{\sigma_F r_B^{\;4} + \sigma_B r_F^{\;4}}}%
      {\left| \sigma_F r_B^{\;2} - \sigma_B r_F^{\;2} \right|} 
 \frac{1}{\sqrt{\epsilon' \int \! dt {\cal L}}} \ , 
\end{eqnarray}
\end{subequations}
where $\epsilon'$ is the detection efficiency for this observation.  
 Since these relative errors are again proportional to 
$1/\sqrt{\epsilon' \int \! dt {\cal L}}$, we introduce the 
nominal relative errors with 1 fb$^{-1}$ of $\epsilon' \int \! dt {\cal L}$, 
$\hat{\cal E}(\sigma^{RR})$ and $\hat{\cal E}(\sigma^{LL})$ 
which are the special values of ${\cal E}(\sigma^{RR})$ and 
${\cal E}(\sigma^{LL})$, respectively, with 1 fb$^{-1}$ of 
$\epsilon' \int \! dt {\cal L}$.  
 Their estimated values are listed in Table~\ref{fberror}.  
 According to the table, we find that $\sigma^{RR}$ is measurable 
within the accuracy of 10{\%} if one accumulates 5.6 fb$^{-1}$ of 
$\epsilon' \int \! dt {\cal L}$, while the statistical error exceeds 
35{\%} for $\sigma^{LL}$ with the same luminosity.  
 To achieve a 10{\%} accuracy for $\sigma^{LL}$, the desired 
$\epsilon' \int \! dt {\cal L}$ is 68---490 fb$^{-1}$, that may be 
realized by a future technology of `an ultimately high luminosity' 
[\ref{ultimate}].  

 The statistical significances of the deviation of $\sigma^{RR}$ and 
$\sigma^{LL}$ from the ones without any Higgs resonances are again 
defined similarly to eq.~(\ref{signif}), and these values with 1 
fb$^{-1}$, $\hat{S}^{RR}$ and $\hat{S}^{LL}$, are also listed in 
Table~\ref{fberror}.  
 As was mentioned in the subsection~\ref{S4.4}, the proposed 
technique to determine CP-parity is not applicable efficiently 
for the multi-Higgs bosons almost degenerated in the cases 
of our numerical estimates.  
 Table~\ref{fberror} shows that the technique requires thousands 
fb$^{-1}$ of the luminosity even in the $\tan\beta$ = 3 case.

\section{Conclusions}
\label{S5}

 We have presented in this paper the effects of heavy CP-even and 
CP-odd Higgs bosons on the production cross section of the process 
$\gamma\gamma$ $\rightarrow t\overline{t}$ at the energy around the 
mass poles of the Higgs bosons.  
 It has been found that the interference between $H$ and $A$ with 
the small mass gap, as well as the ones between Higgs bosons and 
the continuum, contributes to the cross section, if the photon beams 
are polarized and if we observe the helicity of the top quarks.  
 It has been demonstrated in the framework of the MSSM that the $H$ 
and $A$ contributions can be sizable at future $\gamma\gamma$ 
colliders for small value of $\tan\beta$.  
 The methods to evaluate the cross sections, to detect the Higgs 
bosons and to measure the CP-parity of the Higgs boson have also been 
presented.  
 The statistical significances of detecting the Higgs signals and 
measuring the Higgs CP-parity have been evaluated.  
 It has been found that the effective cross section can be measured 
within 10{\%} of statistical accuracy with 1 fb$^{-1}$ of the 
integrated luminosity multiplied by the detection efficiency.  
 The existence of the Higgs signals with $m_A$ = 400 GeV over the 
continuum cross section can be verified in the two-sigma level with 
1.6 fb$^{-1}$ for the $\tan\beta$ = 3 case and 40 fb$^{-1}$ for the 
$\tan\beta$ = 7 case.

\section*{Acknowledgements}
\label{ack}

 This work is supported in part by the Grant-in-Aid for
Scientific Research (No.~11640262) and the Grant-in-Aid
for Scientific Research on Priority Areas (No.~11127205)
from the Ministry of Education, Science and Culture, Japan.  

 The authors would like to thank K.~Hagiwara for valuable discussions 
and reading the manuscript.  
 The authors would also like to thank T.~Kon, T.~Ohgaki, Y.~Okada, 
T.~Tauchi, T.~Takahashi for useful discussions.

\newpage
\section*{References}

\small

\begin{enumerate}
\item
 I.F.~Ginzburg, G.L.~Serbo and V.I.~Tel'nov, {\em Pis'ma Zh.\ Eksp.\ 
 Teor.\ Fiz.}, {\bf 34} (1981) 514 [{\em JETP Lett.}, {\bf 34} (1982) 
 491]; {\em Nucl.\ Instrum.\ Methods}, {\bf 205} (1083) 47.  
\label{ggoriginal}
\item 
 As a summary report, see {\em `$\gamma\gamma$ Collider as an Option 
 of JLC'}, ed.\ I.~Watanabe {\it et al.}, KEK Report 97-17, KEK, March 
 1998.  
\label{ggcol}
\item
 T.~Ohgaki, T.~Takahashi and I.~Watanabe, {\em Phys.\ Rev.}, {\bf D56} 
 (1997) 1723.  
\label{OTW}
\item
 G.~V.~Jikia, {\em Nucl.\ Phys.}, {\bf B405} (1993) 24.  \\
 M.S.~Berger, {\em Phys.\ Rev.}, {\bf D48} (1993) 5121.  
\label{Jikia}
\item
 E.~Asakawa, to be published in the proceedings of {\em ``4th 
 International Workshop on Linear Colliders (LCWS99)"}, Sitges, 
 Barcelona, Spain, 28 Apr. -- 5 May, 1999.  \\
 I.~Watanabe, {\em ibid.}   
\label{asa}
\item
 K.~Hagiwara, H.~Murayama and I.~Watanabe, {\em Nucl.\ Phys.}\ 
 {\bf B367} (1991) 257.  
\label{HMW}
\item
 C.N.~Yang, {\em Phys.\ Rev.}\ {\bf 77} (1950) 242.  \\
 B. Grazadkowski and J.F.~Gunion, {\em Phys.\ Lett.}\ {\bf B291} (1992) 
 361.  \\
 M.~Kr{\"a}mer, J.~K{\"u}hn, M.L.~Stong and P.M.~Zerwas, {\em Z.\ 
 Phys.}\ {\bf C64} (1994) 21.  \\
 J.F.~Gunion and J.G.~Kelly, {\em Phys.\ Lett.}\ {\bf B333} (1994) 
 110.  \\
 S.Y.~Choi and K.~Hagiwara, {\em Phys.\ Lett.}, {\bf B359} (1995) 
 369.  
\label{linearHA}
\item
 See, J.F.~Gunion, H.E.~Haber, G.~Kane and S.~Dawson, {\em `Higgs 
 Hunter's Guide'}, Addison-Wesley Publishing Company, (1990), 
 and references therein.  
\label{Hunter}
\item 
 P.~McNamara, The LEP Higgs Working Group, LEPC meeting, 7 Sept., 
 1999; \\
 \verb+http://www.cern.ch/LEPHIGGS/Welcome.html+.  
\label{mlimit}
\item
 A.~Djouadi, J.~Kalinowski and M.~Spira, {\em Comput.\ Phys.\ 
 Commun.\ Res.}, {\bf 108} (1998) 56.  
\label{HDECAY}
\item
 P.~Chen, G.~Horton-Smith, T.~Ohgaki, A.W.~Weidemann and K.~Yokoya, 
 {\em Nucl.\ Instrum. Methods}, {\bf A335} (1995) 107.  \\
 T.~Ohgaki and T.~Takahashi, {\em Nucl.\ Instrum. Methods}, {\bf A373} 
 (1996) 185.  \\
 {\em `CAIN 2.1b'}, K.~Yokoya, see the CAIN home page, \\
 \verb+http://www-acc-theory.kek.jp/members/cain/default.html+.  
\label{CAIN}
\item
 I.F.~Ginzburg, G.L.~Kotkin, S.L.~Panfil, V.G.~Serbo and V.I.~Telnov, 
 {\em Nucl.\ Instrum.\ Methods Phys.\ Res.}, {\bf 219} (1984) 5.  \\
 V.I.~Telnov, {\em Nucl.\ Instrum.\ Methods Phys.\ Res.}, {\bf A294} 
 (1990) 72.  \\
 D.L.~Borden, D.A.~Bauer and D.O.~Caldwell, SLAC preprint, 
 SLAC-PUB-5715, 1992.  
\label{compton}
\item
 {\em JLC-I}, KEK-Report 92-16 (1992).  
\label{JLC}
\item
 V.~Telnov, {\em Nucl.\ Instrum.\ Methods Phys.\ Res.}, {\bf A355} 
(1995) 3.  
\label{ultimate}
\end{enumerate}

\normalsize

\newpage
\section*{Tables}

\begin{table}[h]
\begin{center}
\caption[Tree helicity amplitudes]{\small 
 The tree helicity amplitudes of $\gamma\gamma$ $\rightarrow 
t\overline{t}$.  
 The overall factor $4 \pi \alpha_{\mbox{\tiny QED}} Q_t^2$ is 
omitted in the table.  }
\vspace{7mm}
\begin{tabular}{|c||c|c|c|c|}
\hline
 & $t_R \overline{t}_R$ & $t_R \overline{t}_L$ 
 & $t_L \overline{t}_R$ & $t_L \overline{t}_L$ \\
\hline \hline
$\gamma_+ \gamma_+$ & 
$- \frac{2 m_t}{E_t} \frac{1+\beta_t}{1-\beta_t^2\cos^2\theta_t}$ & 
$0$ & $0$ & 
$- \frac{2 m_t}{E_t} \frac{1-\beta_t}{1-\beta_t^2\cos^2\theta_t}$ \\
\hline
$\gamma_+ \gamma_-$ & 
$+ \frac{2 m_t}{E_t} 
\frac{\beta_t \sin^2\theta_t}{1-\beta_t^2\cos^2\theta_t}$ & 
$+ 2 \beta_t 
\frac{\sin\theta_t (1+\cos\theta_t)}{1-\beta_t^2\cos^2\theta_t}$ & 
$- 2 \beta_t 
\frac{\sin\theta_t (1-\cos\theta_t)}{1-\beta_t^2\cos^2\theta_t}$ & 
$- \frac{2 m_t}{E_t} 
\frac{\beta_t \sin^2\theta_t}{1-\beta_t^2\cos^2\theta_t}$ \\
\hline
$\gamma_- \gamma_+$ & 
$+ \frac{2 m_t}{E_t} 
\frac{\beta_t \sin^2\theta_t}{1-\beta_t^2\cos^2\theta_t}$ & 
$- 2 \beta_t 
\frac{\sin\theta_t (1-\cos\theta_t)}{1-\beta_t^2\cos^2\theta_t}$ & 
$+ 2 \beta_t 
\frac{\sin\theta_t (1+\cos\theta_t)}{1-\beta_t^2\cos^2\theta_t}$ & 
$- \frac{2 m_t}{E_t} 
\frac{\beta_t \sin^2\theta_t}{1-\beta_t^2\cos^2\theta_t}$ \\
\hline
$\gamma_- \gamma_-$ & 
$+ \frac{2 m_t}{E_t} \frac{1-\beta_t}{1-\beta_t^2\cos^2\theta_t}$ & 
$0$ & $0$ & 
$+ \frac{2 m_t}{E_t} \frac{1+\beta_t}{1-\beta_t^2\cos^2\theta_t}$ \\
\hline
\end{tabular}
\label{tree}
\end{center}

\vspace{4cm}

\begin{center}
\caption[Masses, widths and branching ratios of $H$ and $A$]{\small 
 The masses, the total decay widths, the two-photon decay branching 
ratios and the $t\overline{t}$ branching ratios of $H$ and $A$ 
adopted in the numerical simulations.  }
\vspace{7mm}
\begin{tabular}{|c||c|c|c|c|}
\hline
 $\tan\beta$ & $m_H$ & $\Gamma_H$ & 
               $Br(H \rightarrow \gamma\gamma)$ & 
               $Br(H \rightarrow t\overline{t})$ \\
 & \FS (GeV) & \FS (GeV) & $10^{-5}$ & \\
\hline 
3.0 & 403.79 & 0.79 & 0.99 & 0.742 \\
7.0 & 400.71 & 0.50 & 0.30 & 0.207 \\
\hline \hline
 $\tan\beta$ & $m_A$ & $\Gamma_A$ & 
               $Br(A \rightarrow \gamma\gamma)$ & 
               $Br(A \rightarrow t\overline{t})$ \\
 & \FS (GeV) & \FS (GeV) & $10^{-5}$ & \\
\hline
3.0 & 400.00 & 1.75 & 1.53 & 0.946 \\
7.0 & 400.00 & 0.67 & 0.79 & 0.452 \\
\hline 
\end{tabular}
\label{masses}
\end{center}
\end{table}

\begin{table}[p]
\caption[Signs of Interferences]{\small 
 A rough scheme of the $\sqrt{s}_{\gamma\gamma}$ dependence of the 
signs of the interference terms between ${\cal M}_{\mbox{\tiny tree}}$ 
and ${\cal M}_{H,A}$.  
 The actual switching of the sign occurs at a slightly different 
energy from the mass of the Higgs.  }
\vspace{0.4cm}
\begin{center}
\begin{tabular}{|c|c|c|c|}
\hline
 &
$\sqrt{s}_{\gamma\gamma}< m_A$&
$m_A<\sqrt{s}_{\gamma\gamma}< m_H$&
$m_H<\sqrt{s}_{\gamma\gamma}$
\\
\hline \hline 
{\large $\Delta \sigma_{\mbox{\tiny $A$-tree}}^{++RR}$} & 
$+$ & $-$ & $-$ 
\\
{\large $\Delta \sigma_{\mbox{\tiny $H$-tree}}^{++RR}$} & 
$+$ & $+$ & $-$ 
\\
\hline 
{\large $\Delta \sigma_{\mbox{\tiny $A$-tree}}^{++LL}$} & 
$+$ & $-$ & $-$ 
\\
{\large $\Delta \sigma_{\mbox{\tiny $H$-tree}}^{++LL}$} & 
$-$ & $-$ & $+$ 
\\ \hline
\end{tabular}
\label{tab:exp}
\end{center}

\vspace{2cm}

\begin{center}
\caption[Effective cross section]{\small 
 The effective cross sections and these relative errors 
with 1 fb$^{-1}$ 
for nine values of the collider energy at $\tan\beta$ = 3 and 7.  
 The effective continuum cross section and the statistical 
significance with 1 fb$^{-1}$ of the deviation of the cross 
section with the 
Higgs resonances from the continuum are also listed.  
 The sign of $\hat{\cal S}$ indicates the 
sign of the deviation.  }
\vspace{7mm}
\begin{tabular}{|c|c||cc|cc|}
\hline
$\tan\beta$ & $\sqrt{s}_{ee}$ & 
  $\sigma$ & $\hat{\cal E}(\sigma)$ & 
  $\sigma_{\mbox{\tiny tree}}$ & $\hat{\cal S}$ \\
 & \FS (GeV) & \FS (fb) &  & \FS (fb) &  \\
\hline \hline
3.0&     480&   121.1&   0.091&   110.1&   $+1.00$ \\
   &     490&   141.4&   0.084&   122.6&   $+1.58$ \\
   &     500&   145.9&   0.083&   131.8&   $+1.17$ \\
   &     510&   140.9&   0.084&   138.6&   $+0.19$ \\
   &     520&   135.6&   0.086&   143.6&   $-0.69$ \\
   &     530&   134.7&   0.086&   147.3&   $-1.09$ \\
   &     540&   137.5&   0.085&   149.9&   $-1.06$ \\
   &     550&   141.5&   0.084&   151.5&   $-0.84$ \\
   &     560&   144.6&   0.083&   152.4&   $-0.65$ \\
\hline
7.0&     480&   112.0&   0.094&   110.1&   $+0.18$ \\
   &     490&   125.0&   0.089&   122.6&   $+0.21$ \\
   &     500&   131.8&   0.087&   131.8&   $+0.00$ \\
   &     510&   136.1&   0.086&   138.6&   $-0.21$ \\
   &     520&   140.0&   0.085&   143.6&   $-0.30$ \\
   &     530&   143.9&   0.083&   147.3&   $-0.29$ \\
   &     540&   147.3&   0.082&   149.9&   $-0.22$ \\
   &     550&   149.7&   0.082&   151.5&   $-0.15$ \\
   &     560&   151.1&   0.081&   152.4&   $-0.11$ \\
\hline
\end{tabular}
\label{total}
\end{center}
\end{table}

\begin{table}[p]
\begin{center}
\caption[Forward and Backward Methods]{\small 
 The effective cross sections $\sigma^{RR}$, $\sigma^{LL}$, 
$\sigma^{RL+LR}$, $\sigma_F$, $\sigma_B$.  
 The decay fractions in forward $r_F$ and backward $r_B$ of the 
right-handed top quark are assumed to be 0.40 and 0.60, 
respectively.  }
\vspace{7mm}
\begin{tabular}{|c|c||rrc|cc|}
\hline
$\tan\beta$ & $\sqrt{s}_{ee}$ & 
 $\sigma^{RR}$ & $\sigma^{LL}$ & $\sigma^{RL+LR}$ & 
 $\sigma_F$  & $\sigma_B$  \\
 & \FS (GeV) & \FS (fb) & \FS (fb) & \FS (fb) & 
   \FS (fb)  & \FS (fb) \\
\hline \hline
3.0&     480&    99.1&    21.9&     0.1&    23.7&     39.4 \\
   &     490&   117.0&    24.2&     0.2&    27.4&     46.2 \\
   &     500&   120.4&    25.2&     0.3&    28.3&     47.6 \\
   &     510&   117.6&    22.9&     0.4&    27.1&     46.3 \\
   &     520&   116.4&    18.7&     0.5&    25.4&     45.2 \\
   &     530&   119.5&    14.6&     0.6&    24.4&     45.7 \\
   &     540&   125.2&    11.6&     0.6&    24.2&     47.3 \\
   &     550&   130.9&     9.8&     0.7&    24.5&     49.1 \\
   &     560&   135.1&     8.7&     0.8&    24.8&     50.5 \\
\hline
7.0&     480&    92.1&    19.8&     0.1&    21.8&     36.5 \\
   &     490&   105.9&    18.9&     0.2&    23.7&     41.4 \\
   &     500&   114.3&    17.2&     0.3&    24.4&     44.2 \\
   &     510&   120.4&    15.4&     0.4&    24.8&     46.1 \\
   &     520&   125.9&    13.7&     0.5&    25.1&     47.8 \\
   &     530&   131.1&    12.2&     0.6&    25.4&     49.5 \\
   &     540&   135.6&    11.0&     0.6&    25.7&     51.0 \\
   &     550&   139.0&     9.9&     0.7&    25.8&     52.1 \\
   &     560&   141.3&     9.0&     0.8&    25.9&     52.8 \\
\hline
\end{tabular}
\label{fb}
\end{center}
\end{table}

\vfil

\begin{table}[p]
\begin{center}
\caption[Errors of Forward and Backward Methods]{\small 
 The relative errors of the effective cross sections, 
$\hat{\cal E}(\sigma^{RR})$ and $\hat{\cal E}(\sigma^{LL})$, 
and these statistical significances of the deviation of the 
cross section with the Higgs resonances from the continuum, 
$\hat{\cal S}^{RR}$ and $\hat{\cal S}^{LL}$.  
 Here we assumed 1 fb$^{-1}$ of the luminosity multiplied by 
the detection efficiency.  
 The sign of the statistical significances indicates the 
sign of the deviation.  }
\vspace{7mm}
\begin{tabular}{|c|c||cc|cc|}
\hline
$\tan\beta$ & $\sqrt{s}_{ee}$ & 
 $\hat{\cal E}(\sigma^{RR})$ & $\hat{\cal E}(\sigma^{LL})$ & 
 $\hat{\cal S}^{RR}$ & $\hat{\cal S}^{LL}$ \\
 & \FS (GeV) & & & & \\
\hline \hline
3.0&     480&    0.23&    0.87&   $+0.38$ &   $+0.13$ \\
   &     490&    0.21&    0.85&   $+0.53$ &   $+0.29$ \\
   &     500&    0.21&    0.83&   $+0.23$ &   $+0.40$ \\
   &     510&    0.21&    0.89&   $-0.22$ &   $+0.38$ \\
   &     520&    0.21&    1.05&   $-0.53$ &   $+0.25$ \\
   &     530&    0.20&    1.33&   $-0.61$ &   $+0.11$ \\
   &     540&    0.20&    1.66&   $-0.52$ &   $+0.02$ \\
   &     550&    0.19&    1.96&   $-0.39$ &   $-0.01$ \\
   &     560&    0.19&    2.21&   $-0.29$ &   $-0.02$ \\
\hline
7.0&     480&    0.24&    0.93&   $+0.07$ &   $+0.02$ \\
   &     490&    0.22&    1.02&   $+0.08$ &   $+0.03$ \\
   &     500&    0.21&    1.13&   $-0.02$ &   $+0.02$ \\
   &     510&    0.20&    1.28&   $-0.10$ &   $+0.00$ \\
   &     520&    0.20&    1.44&   $-0.14$ &   $-0.01$ \\
   &     530&    0.19&    1.62&   $-0.13$ &   $-0.01$ \\
   &     540&    0.19&    1.80&   $-0.10$ &   $-0.01$ \\
   &     550&    0.19&    1.99&   $-0.07$ &   $-0.01$ \\
   &     560&    0.18&    2.20&   $-0.05$ &   $-0.00$ \\
\hline
\end{tabular}
\label{fberror}
\end{center}
\end{table}

\vfil

\newpage

\newpage


\newpage

\newpage

\vfil

\end{document}